\documentclass[10pt, journal, a4paper, final, twoside, nofonttune]{IEEEtran}
\usepackage{epsfig,subfigure,float}
\usepackage[russian, english]{babel}

\begin{document}
\title{Unconventional research in USSR and Russia: \\short overview}
\author{Serge~Kernbach$^1$\\[3mm]
\thanks{$^1$Cybertronica Research, Research Center of Advanced Robotics and Environmental Science, Melunerstr. 40, 70569 Stuttgart, Germany, \emph{serge.kernbach@cybertronica.co}}
\thanks{$^2$V.1.4. Russian version of this work is submitted to the \emph{International Journal of Unconventional Science}.}
}

\maketitle
\thispagestyle{empty}

\begin{abstract}
This work briefly surveys unconventional research in Russia from the end of the 19th until the beginning of the 21th centuries in areas related to generation and detection of a 'high-penetrating' emission of non-biological origin. The overview is based on open scientific and journalistic materials. The unique character of this research and its history, originating from governmental programs of the USSR, is shown. Relations to modern studies on biological effects of weak electromagnetic emission, several areas of bioinformatics and theories of physical vacuum are discussed.
\end{abstract}

\section{Introduction}
\Rus

$^2$In the USSR and Russia, from about 1921 up until now, many different research activities, denoted today as unconventional, have been conducted. For instance, these are related to the impact of weak and strong electromagnetic emission on biological objects, quantum entanglement in macroscopic systems, non-local signal transmission based on the Aharonov-Bohm effect, 'human operator' phenomena, and others. This research, especially after the 80's, was very broad -- it dealt with several boundary areas of psychology on one side and touched theoretical physics on other side. Since USSR had in fact no unsupported-by-government research, unlike Europe and USA, where such research can be supported by private funds, all these activities can be interpreted as government programs. Single scientists such as A.L.Chigevskiy (А.Л.Чижевский), N.A.Kozyrev (Н.А.Козырев) or N.I.Kobosev (Н.И.Кобозев), who worked outside state programs, encountered substantial difficulties and were almost unknown in their time.

Several such governmental programs are not officially published up to now. For instance, documents on experiments performed in OGPU and NKVD -- even 80 years after -- still remain classified. Information about these works originates mainly from indirect sources, such as interviews of participating persons, solitary scientific and popular publications. From popular sources, works \cite{Vinokurov93,Vinokurov05,Shichkin99,Andrees04,Kolpadi98,Greg12} can be mentioned. More scientifically based works are \cite{Dubrov89},\cite{Naumov93}, \cite{Dulnev98}, \cite{PhenomenD91}, \cite{parapsi}, \cite{Bogachhin03}, from last years -- \cite{Melnik10}, \cite{Zhigalov11}. Works \cite{May92RUS}, \cite{May93RUS}, \cite{Maire11}, \cite{Collins86}, \cite{Menzel12}, \cite{Mannherz12}, \cite{Ostrander70} are foreign sources about the state of unconventional research in the USSR and Russia in this area. Unfortunately, there are also a number of unserious publications. In this situation it is difficult to survey this research unambiguously.

In this work we attempt, based on a large number of open publications, to estimate the boundary of unconventional research in the USSR and Russia. We demonstrate that these works are historically concentrated within three large areas: (1) long-distance biological signal transmission, inc. plants, animal and humans, mind-matter phenomena, different ESP effects and similar topics; (2) non-ionizing, in particular electromagnetic, emissions from human and its impact on human physiology and more generally on different biological systems; (3) phenomena related to the generation and detection of a 'high-penetrating' emission from biological and non-biological origin. A large part of this article is focused primarily on (3) in the non-biological context. In the terminology of this paper, the areas (2) and (3) are considered as the areas of unconventional research. For an in-depth overview, especially in the 'classical' area (1), we can suggest other works, e.g. E.K.Haumov (Э.К.Наумов) and colleagues \cite{Naumov93}, A.P.Dubrov and V.N.Pushkin (А.П.Дубров, В.Н.Пушкин) \cite{Dubrov89} and others. Since unconventional research in the USSR and Russia has a cyclic character, we distinguish three following periods: from 1917 to 1937, from 1955 to 1980, from 1980 to 2003. These periods are different from the viewpoint of their character, obtained results and position of government. To create a full picture of the situation, we shortly demonstrate the state before 1917 and after 2003.

This work is prepared within the presentation on the same topic in the institute for frontier areas of psychology and mental health (IGPP institute) in Freiburg, Germany, in October 2013. We pointed out historical roots of the unconventional research, caused primarily by a specific Soviet funding strategy, and its evolution, which developed differently from similar Western works. The positive potential of obtained results, e.g. for a signal transmission on long and super-long distances, metallurgy, biological and biophysical areas, as well as a possibility of their unethical applications, were discussed.

This paper has the following structure. In Sec.~\ref{sec:background} several prerequisites of unconventional research in the USSR and Russia are shown. Sec.~\ref{sec:1917}, \ref{sec:19171937}, \ref{sec:19551980}, \ref{sec:19802003} are devoted to corresponding historical periods. In Sec. \ref{sec:dev2003} we shortly survey the current state of art. Finally, this work is summarized in Sec.~\ref{sec:conclusion}.

\section{Understanding the background of unconventional research}
\label{sec:background}

For a proper understanding of the situation in Russia and the USSR in the field of unconventional research, first of all it needs to note the positive perception of unusual and supernatural phenomena among the general population. In folklore, mythical creatures such as 'Baba Yaga', 'Koshei the Deathless', 'Zmey Gorynych', the Princess Frog, brownies and others, as well as the popular belief in e.g. 'evil eye' -- are an integral part of fairy tales, novels, films and are in cultural, publicity and everyday use. This cultural specificity, in the terminology of positive 'operator effect' \cite{BLASBAND00}, \cite{Dunne88}, \cite{Dunne95} was also emphasized by several western researchers \cite{May92RUS}.

The contacting of psychics within regime -- G.J.Rasputin and the last Tsar' family \cite{Mannherz12}, W.Messing and J.W.Stalin \cite{Nepomnjshy99}, E.Y.Davitashvili ('Juna') and L.I.Brezhnev \cite{Mularov99} or a group of psychics at B.N.Yeltsin \cite{Interfax07}, \cite{Kusina07} -- were present in the days of the Russian Empire and the Soviet Union. Due to such contacts, many of the phenomena demonstrated by these people are scientifically explored by the Soviet Academy of Science. Despite the fact that the Soviet Union conducted a large amount of state-funded unconventional research, the official position to several aspects of this research, especially in boundary areas of psychology, was negative. It was conditioned primarily by ideological considerations. Unfortunately, in Russia after 1991, there is the same tendency, however originating not from the political, but from the academic community. Thus, there is an extremely polarized response of different government agencies, where there are a lot of supporters and opponents of these programs at various levels of hierarchy. Depending on the involvement into governmental programs, researchers can receive support and funding, and vice versa lose their jobs (and freedom at the time of the NKVD) for being engaged in 'alternative technologies'.

So-called alternative or traditional medicine is widespread: 'according to data published by the Research Center of the Academy of Medical Science, 80 percent of patients are addressed by the healers and sorcerers' \cite{Prilepina07}. In Soviet time 'unlawful practice of medicine' was a criminal offense (Article 221 of the Criminal Code of the RSFSR), in 1997 the concept of 'traditional medicine' was introduced, for which a 'healer diploma' and not a medical license was required. Horoscopes drawn up on a variety of topics -- the so-called 'astro-forecast (astroprognoz)' -- can be found on pages of many newspapers, even central ones such as \emph{Russiyskay Gazeta}. At the making of 'Battle of psychics'\footnote{http://bitvaextrasensov.tnt-online.ru/} (Russian version of 'Psychic Challenge') a long queue of people appeared each time, who consider themselves gifted by various abilities.

It is necessary to note a large appearance of New Age literature after 1991. Almost all publishers have one or several series devoted to this topic. This situation is radically different from the state before 1991. In the USSR, esoteric literature was generally prohibited, and published in small amounts in the form of underground samizdat \cite{Strukova10}. Among the authors, it needs to point out E.P.Blavatsky \cite{Blavataskaj_Doktrina}, especially in the Soviet era, whose Hermetic philosophy had a great influence on researchers.

\section{Before 1917}
\label{sec:1917}

The first official contact between the Russian science and phenomena, denoted today generally as paranormal, occurred in the 70s of the 19th century. The commission, headed by Prof. D.I.Mendeleev (Д.И.Менделеев), investigated the phenomena of Spiritualism \cite{Mendeleev76}. The commission included, among others, Prof. A.M.Butlerov (А.М.Бутлеров) and Prof. N.P.Vagner (Н.П.Вагнер) who believed in the existence of these phenomena. The traces of the work of this Commission can be found in correspondence between the Russian and German scientists and even in the works of H.P.Blavatsky \cite{Blavatskaj06}. The conclusion of the commission in 1876:
\begin{quote}
'Spiritualistic phenomena occur on the unconscious movements or conscious deception, and spiritualistic doctrine is superstition'.
\end{quote}
Despite the Commission denying these phenomena, many questions remained open. Taking into account books and periodic publications of A.M.Butlerov, it can be assumed that further discussion was sharply polarized. This polarity in the acceptance and rejection of research, touching (even indirectly) on any paranormal phenomena, is characteristic for all future Russian and Soviet studies.

\begin{figure}[thp]
\centering
\includegraphics[width=0.49\textwidth]{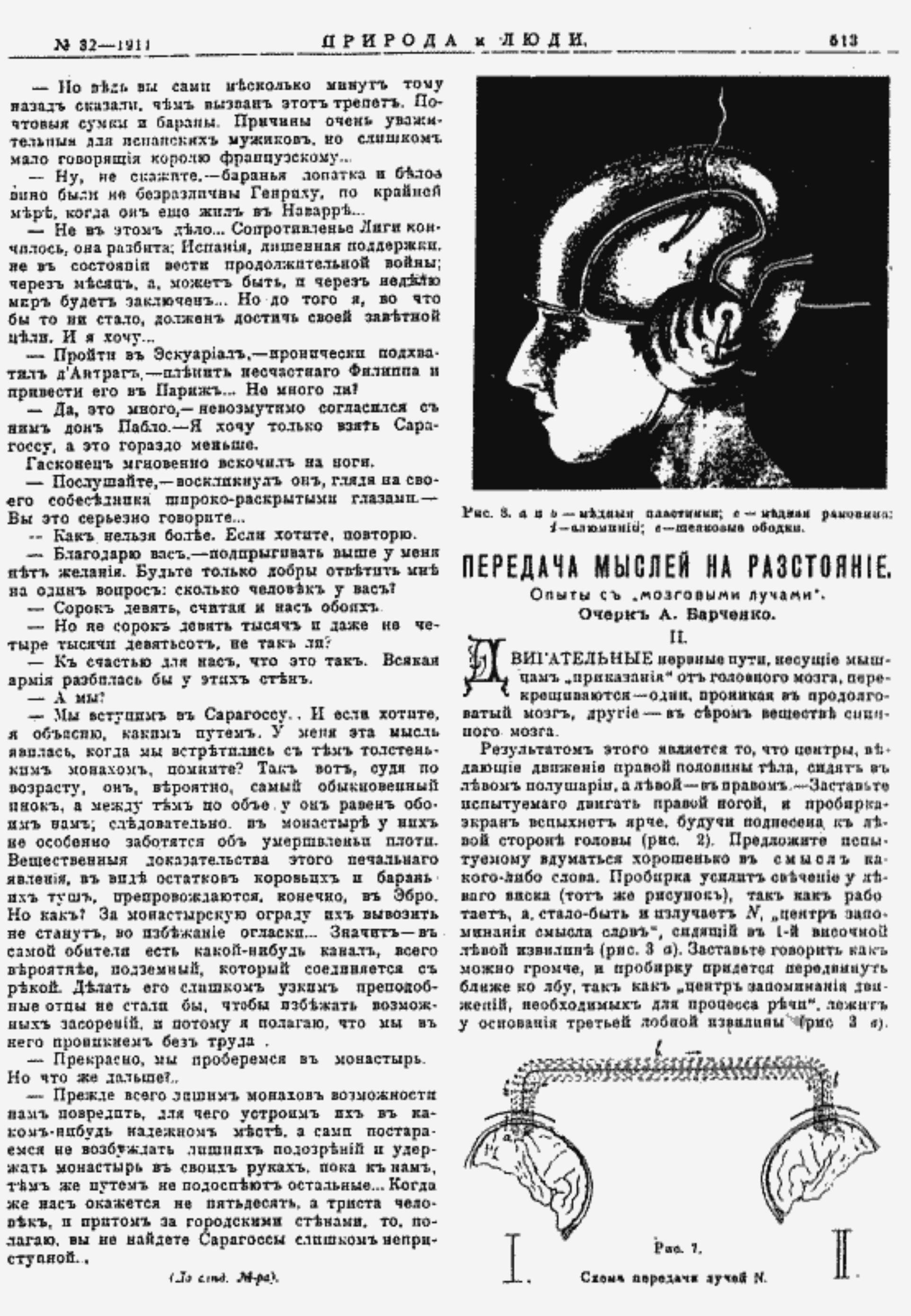}
\caption{\em \small Paper of A.Barchenko (А.Барченко) on 'mental suggestion' and a distant transfer of thoughts, published in 1911. A.Barchenko is known from the Soviet OGPU programs in 1921-1937. \label{fig:Abrams}}
\end{figure}

To explain the phenomenon of mental suggestion, Butlerov hypothesized that the nervous system and the brain are a source of specific radiation. By analogy, the movement of 'nerve currents' in the body is similar to the electric currents in conductors, as shown in Fig.~\ref{fig:Abrams}. Transmission of signals from the brain of one person to another one occurs due to the 'electro-inductive' effect. Despite this assumption sounding naive from today's point of view, this attitude was typical for the late 19th century. It was primarily motivated by developments of radio at that time, where many researchers tried to find 'radio-waves' also in the psychic area. For example, W.Crookes assumed the human brain is capable of sending and receiving some sort of electromagnetic rays of very high frequency. A similar view was shared also by Albert Abrams \cite{Russell97}, whose work led to the emergence of radionics in USA. However, Russian engineering works, despite a similar start in the 19th century, evolved in a different direction as it will be shown later.

Vitalistic theories also had their own development in Russia. For example, the famous psychiatrist Y.L.Ohotrovich (Ю.Л.Охотрович), since 1867 developed magnetic hypnotic therapy based on the theory of animal magnetism (Mesterism) \cite{Moroz91}. Following Ohotrovich, all living organisms emit a 'particular magnetic field' (in the Ohotrovich' terminology -- Hard-rays), which is of organic origin. In 1910-1912 he lectured on this subject and even was awarded by the Paris Academy of Sciences. Ohatrovich's great merit is the pursuit of interdisciplinary works on studying these phenomena.
\begin{quote}
'It should be mentioned that Y.Ohorovich has devoted many years to study the phenomena of telepathy and mediumship. After a series of experiments with well-known medium E.Palladino, the scientist came to the conclusion that these are manifestations of an organic energy of the medium, which can be investigated by experimental methods. The result of his research was a five-volume work 'mediumistic phenomena'. In his book 'The Secret Knowledge in Egypt' (1894), which complements even the modern Egyptology, Y.Ohorovich argues that paranormal phenomena, being physical, consist in the transmission of a low-level energy, which are however measurable by technical means' \cite{Moroz91}.
\end{quote}

It can be assumed that between the Commission's report in 1876 and the revolution of 1917, there was an interest in the study of mental phenomena by means of physiological research. It was driven by a scientific community not only in Russia but also abroad. For example, in 1890 the Russian society of experimental psychology formed a commission to study the phenomenon of 'mind reading'. In 1907 it published the works of the first All-Russian Congress of Spiritualists. In 1907, by decree of Emperor Nicholas II the St. Petersburg Research Institute of neuropsychiatrics was founded \cite{Akimienko00}. Its founder was V.M.Bechterev (В.М.Бехтерев). St. Petersburg' institute was one of the main centers, where unconventional studies were later conducted. Bechterev's works before 1917 are related to various aspects of mental suggestion:
\begin{quote}
'In view of all what has been said, it is impossible not to agree with the fact that hysteria and spoilage largely owe their origin to the side of home life of Russian people. It is obvious that peculiar superstitions and religious beliefs of people give a psychic coloring of the disease state, which is known under the name of spoilage, hysterics and possession. The question of the development of these diseases in our nation is deeply interesting. In this respect, it plays apparently a huge role involuntary suggestion experienced by individuals under different conditions' \cite{Bechterev08}.
\end{quote}
To summarize the period before 1917, it should be noted that a well-known school of 'mental suggestion' was formed in Russia. Here some vitalistic works and original psychological developments can be found. On the technological side, there is a tendency to explain these phenomena with the help of electromagnetism, which is typical for that time. In performed experiments we observe an attempt to measure these mental effects. Due to extensive contacts between Russian and European researchers, there was a similar formulation of problems and experimental methodology. In subsequent periods, the number of scientific communications was dramatically reduced, and the topics of Russian works increasingly took a specific character, which was subsequently referred to as 'the Soviet program'.

\section{The period between 1917 and 1937}
\label{sec:19171937}

The period after 1917 is primarily related to the October Revolution. As we will see, the unconventional research in Russia is characterized by periodical movement from broad support to almost total closing. The first such period began in 1917-1918. Organizations, not supported by the new regime, were closed. This was especially related to masonic and spiritualistic movements, see \cite{Kolpadi98}. The parapsychological journal 'Rebus', published since 1881, was closed \cite{Naumov93}. There appeared a new scientific system, where scientists also played some roles in relevant government agencies. For example, Bechterev was a member of the Academic Council at the Commissariat (Ministry) of Education \cite{Bechterev28}.

It can be assumed that a, more or less, coordinated Soviet program began in 1924, when the Commissar (Minister) of Education A.V.Lunacharskiy (А.В.Луначарский) formed the Russian Committee for Psychical Research at the International Committee of the Psychical Research. Many authors (e.g. \cite{Plexanov04}, \cite{Naumov93}, \cite{Vinokurov93}) point to a program of the USSR's Commissar of Defense in 1932-1937, related to transfer of information in a biological way. These works were conducted in two places: in Leningrad at the Bechterev's Brain Institute, led\footnote{V.M.Bechterev died in 1927.} by Prof. L.L.Vasilyev (Л.Л.Васильев) and in Moscow at the laboratory of biophysics, Academy of Science, led by Prof. P.P.Lazarev (Director of Laboratory) and Prof. S.Y.Turlygin (П.П.Лазарев, С.Я.Турлыгин). The Biophysics laboratory was asked to investigate the physical nature of telepathy. For instance, in Moscow's laboratory the first results of biological emission from humans were obtained. In Leningrad, the Brain Institute was requested to perform more psychologically oriented works, such as transferring visual images and remotely influencing the percipient. Both research organizations did not know about the works of each other \cite{Vinokurov93}.

\begin{figure}[thp]
\centering
\includegraphics[width=0.47\textwidth]{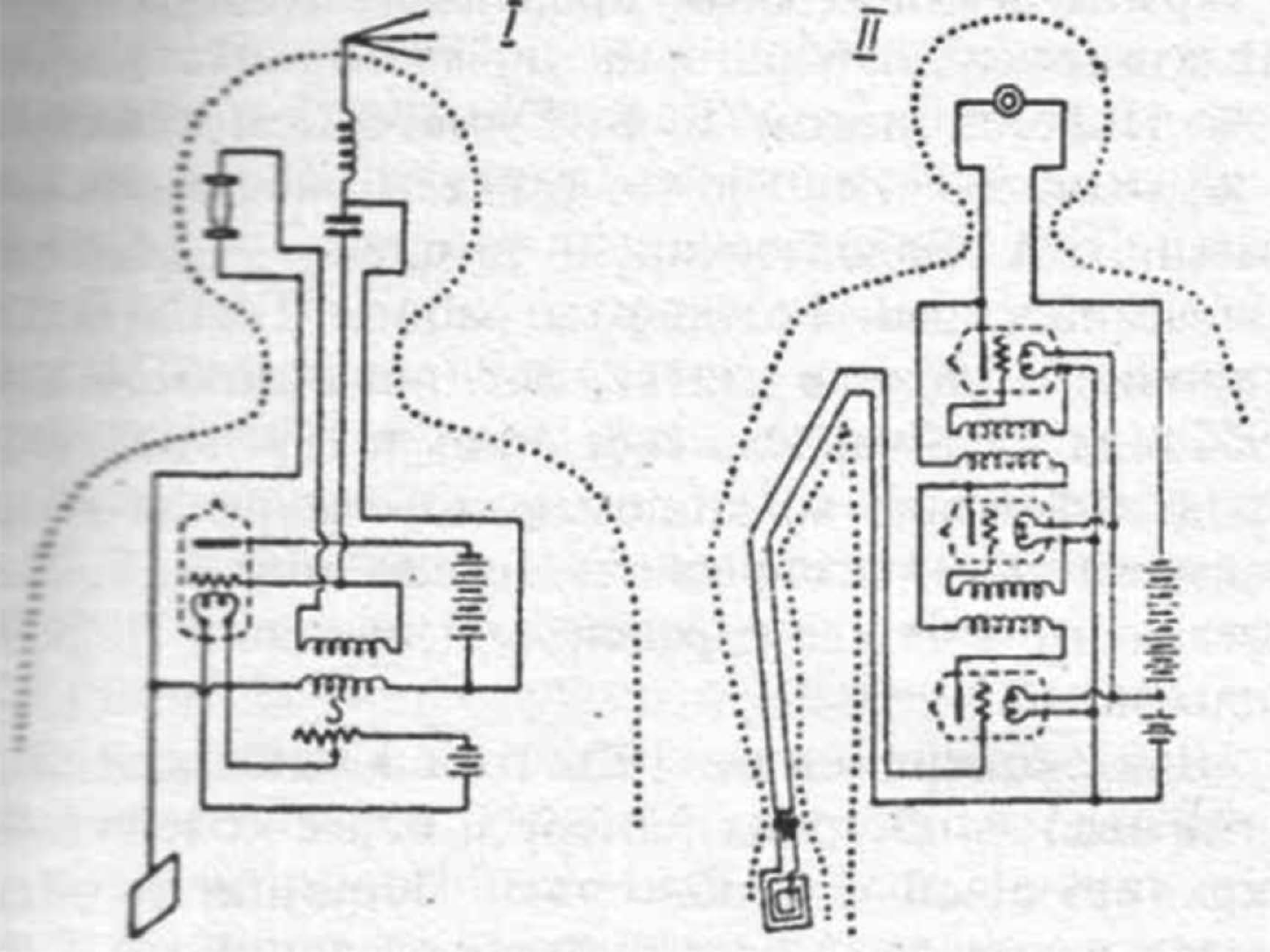}
\caption{\em \small The original scheme of transmitting and receiving bio-circuitry of the human nervous system, image by B.B.Kazhinskiy \protect\cite{Kajinski63}.\label{fig:kaginskij}}
\end{figure}

The book by B.B.Kazhinskiy (Б.Б.Кажинский) \cite{Kajinski63} points to works of A.V.Leontovych (А.В.Леонтович), L.L.Vasiliev, V.M.Bechterev and P.P.Lazarev, which are related to electrophysiology and were made between 1916 and 1921. Thus, in the 20s there were various groups of researchers, both in Leningrad and Moscow. The fundamental works of B.B.Kazhinsky, A.V.Leontovich, L.L.Vasiliev and V.M.Bechterev focus on the effect of suggestion \cite{Bechterev08} and signal transmission in a biological way, see Fig.~\ref{fig:kaginskij} (for example experiments with animals performed by Lev Durov (Л.Дуров) \cite{Durov24}).

It was already clear in 1927, that electromagnetism does not explain the nature of telepathy:
\begin{quote}
'Chambers (Faraday cage) were made of metal. The subjects were first placed in the chambers, then the experiments were conducted outside the chambers. There was no difference! The phenomenon of telepathy manifested equally in the chamber and outside it! It turned out that the iron walls of the chambers were not a barrier to telepathic radiation? Well, then it is not radio waves...' \cite{Kolpadi98}.
\end{quote}
Since these experiments, the 'telepathic radiation' is denoted as 'high-penetrating' and 'non-electromagnetic'. However, the works with EM emission have shown that mechanisms of higher nervous activity can be affected by technical means, in particular by microwave radiation. In this context, the effects discovered by B.G.Michaylovskiy (Б.Г.Михайловский) can be mentioned. It is about impact of medium/short EM waves modulated by low-frequency signals on separate areas of the brain, which are responsible for emotional state and functionality of different organs \cite{Kolpadi98}. Turlygin in the introduction to his work \cite{Tyrlygin42} described the biological effects of electromagnetic radiation:
\begin{quote}
'Various biological effects can be explained by unequal absorption of microwave energy by different tissues of the body, as well as by an excitation of some parts of a nervous system ... This method is new and apparently powerful enough to estimate this absorption of electromagnetic energy from both qualitative and quantitative sides. Indeed, if the absorption of electromagnetic energy takes place in nerves and nerve cells, a negligible portion of the energy absorbed by the nerve is enough to produce a number of secondary phenomena in the body caused by the excitation of a nerve. It is essential that such secondary processes occur due to a comparatively high internal energy of the body (for example, due to the energy of tissues, which are innervated by the 'modified' nerve in our experiments), but not due to a relatively insignificant absorbed external energy by the nerve. Thus, all parts of the electromagnetic spectrum contains some bands, which act differently on biological objects, in particular, on the nervous system of a living organism ... The search for radiation emitted by organisms represents the second way of finding biologically active waves and of clarifying the issues of absorption of the electromagnetic field. If the experiment confirms the existence of an active emitting of electromagnetic waves by a living organism, it is necessary to clarify the biological significance of it, since it is no matter how small is its intensity, it can be one of the strongest factors in the daily life of living organisms' \cite{Tyrlygin42}.
\end{quote}
Thus, Turlygin confirms the orientation and the subject of the Soviet pre-war program. This program, as it is followed from open publications, focused on problems of information transfer and their practical applications for biological objects. It must be said that similar works were carried out in other countries, see e.g. the book of Kazhinskiy \cite{Kajinski63}. For example, the well-known experiments were made by Ferdinando Cazzamalli in Italy for the detection of meter and centimeter EM waves emitted by the brain during increased mental activity \cite{Cazzamalli63}. As it can be seen from the publishers of this book (Research Laboratory of the U.S. Army, fort Balvuar), this research was also pursued in other countries.

The role of special services in shaping the USSR' unconventional programs should be mentioned separately. Apparently, OGPU-NKVD was interested in the possibilities of this technology. As an example, the name of A.Barchenko can be indicated. In 1921 he organized an expedition to the Kola Peninsula on Bechterev's orders \cite{Demin99}. Another Barchenko's expedition to the Crimea was financed by the OGPU, and was related to the chief of special department G.Bokia (Г.Бокия). Some sources mentioned the existence of several special laboratories. For example, one of them - 'neuro-energetic' laboratory at the special department of the OGPU \cite{Kolpadi98} - was initially with the Moscow Power Engineering Institute, after 1934 (1935) - in the building of the Institute of Experimental Medicine. The name of Barchenko's work is very characteristic: 'Introduction to the methodology of experimental influence of volumetric energy field' \cite{Andrees04} (unfortunately the text of the work is not available). Funding of some Barchenko' projects was apparently performed on personal instructions of F.Dzerzhinskiy (Ф.Дзержинский) \cite{Demin99}, \cite{Kolpadi98}. Due to the fact that Barchenko's documents are still classified, we can only guess what was happening in his laboratory.
\begin{quote}
'(Barchenko) undertook repeated attempts to organize an expedition to Tibet, made a trip to the caves of the Crimea, 'bear corners' of the Kostroma region, the Altai, where many occult objects were collected... The scientist was a consultant during investigations of all kinds of healers, shamans, psychics and hypnotists. To test these 'human anomalies' Bokia's department equipped a special 'black room'' \cite{Kolpadi98}.
\end{quote}

\begin{figure}[thp]
\centering
\includegraphics[width=0.4\textwidth]{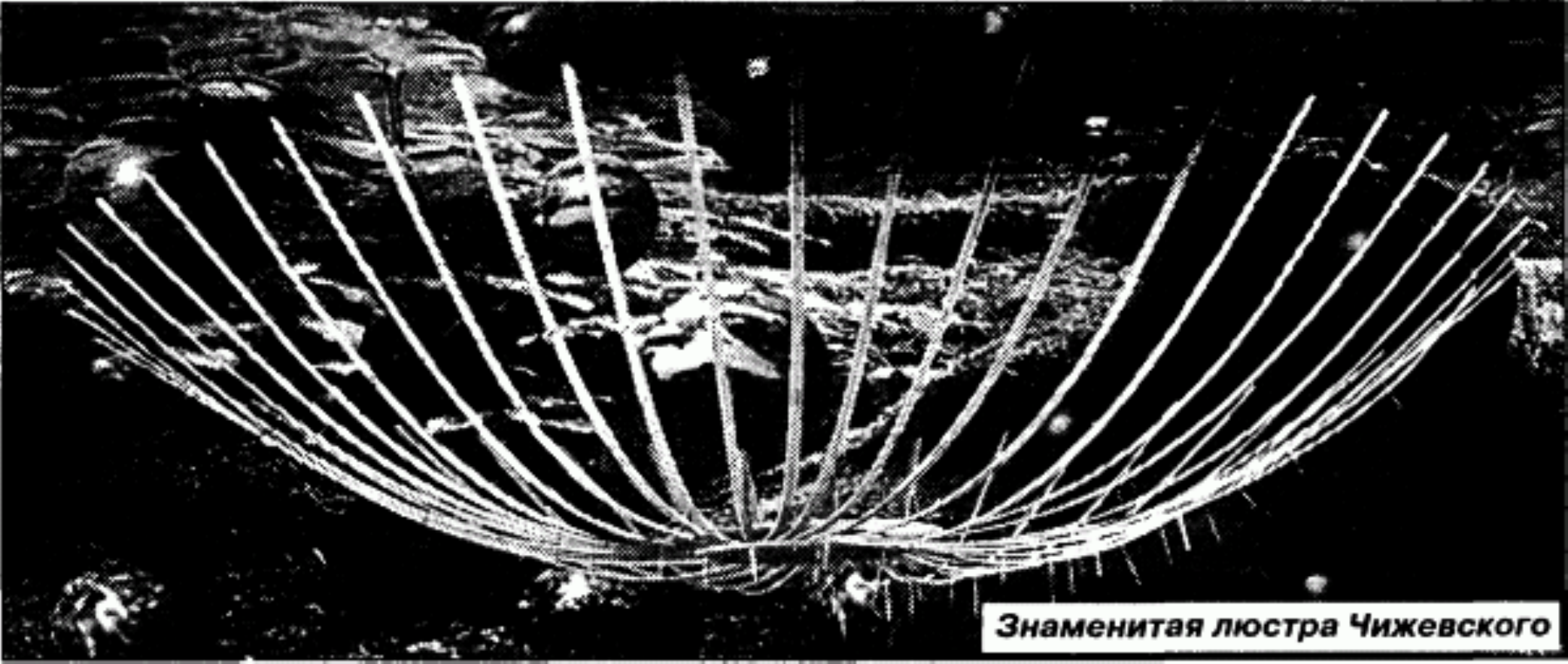}
\caption{\em \small So-called 'Chizhevskiy Chandelier'. A high voltage is applied to this object to create the aeronisation effect.
\label{fig:veinik}}
\end{figure}

The pre-war research was performed within a narrow program, supervised by the state. Single researchers, even if they obtained interesting results in this area, were strongly persecuted. In this context the name of A.L.Chizhevskiy can be mentioned. Despite Chizhevskiy collaborated with Bechterev and Kazhinskiy in Durov's zoo-psychological laboratory \cite{Vinokurov93}, his fate is different. In the years 1930-1936 he expressed an interesting theory about unknown Z-rays in sun emission. These rays are presumably of non-electromagnetic nature, possess a 'high-penetrating' character and interact with biological organisms \cite{Chigevsky73}. The standard test for erythrocyte sedimentation under the influence of 'high-penetrating' emission has also been developed by Chizhevskiy \cite{Chigevsky73}, \cite{Chigevsky80}. He invented so-called 'Chizhevskiy Chandelier' in the 30s, which has an effect of aeronisation, see Fig.~\ref{fig:veinik}. As it revealed later, this device is also a generator of 'high-penetrating' emission \cite{Shkatov10}, whose design is similar to another generator -- the so-called 'Veinik's chronal generator' of the 90s.

Chizhestvskiy's Z-rays can be related to the works of N.P.Myshkin (Н.П.Мышкин), who introduced 'pondemotor forces of the light field' and also have a 'high-penetrating' character \cite{Myshkin06}. In connection with Chizhevskiy, it needs to mention the name of K.E.Ziolkovskiy (К.Э.Циолковской), who was also a supporter of \emph{cosmism} and whose philosophic and ethic works had been classified for a long time after his death \cite{Chigevsky95}.

As indicated by many sources, all these studies and programs were partially or completely collapsed in 1937. Some of the researchers and their families were repressed. Barchenko, Bokia, as well as some members (see e.g. \cite{Shichkin99}) of Roerich's Central-Asian expedition were shot \cite{Kolpadi98}. In 1936 Chizhevskiy was dismissed as 'incompetent' and in 1942 was arrested for 'counter-revolutionary activities' and exiled. All research results and manuscripts were classified, only S.J.Turlygin published abridged results of his research in the field of human microwave radiation in the early 40s \cite{Tyrlygin42}.

Closing programs for almost 20 years indicated the first cycle in the development of unconventional research (and also several more general areas of parapsychology) in the USSR. Due to the tightening of the totalitarian regime, war and ideological differences several western works have been 'overlooked' and remained almost unknown in the USSR, e.g. in the field of radionics, screw effects and instrumental parapsychology. These works also dealt with 'high-penetrating' emissions, but from different points of view. For example, the idea of 'subtle-field resonances', which is characteristic for radionics, has not found its way into the Soviet Union, at least in the 60s and 70s. This is also related to the 'strange non-biological radiation' detected by Chizhevskiy and Myshkin, which was also found in the West, for example in the works of Thomas Hieronymus (including radiation from celestial bodies) \cite{Russell97}, \cite{Hieronymus31} and Victor Schauberger \cite{Rodier99}. Further Soviet works at the end of the 50s and early 60s began with repeating experimental results made prior to 1937. \looseness=-1

\section{The period from 1955 until 1980}
\label{sec:19551980}

Many sources indicate that the post-war Soviet works began with a study of Germany's 'Ahnenerbe' archives removed from the castle 'Altan' \cite{Lavronov00}. We cannot say whether this is true or not. On the one hand, these sources claim the documents have not been worked out in the 60s (only later in 80s-90s). In publications of that period, we cannot detect any significant breakthrough in the field of unconventional research. However, on the other hand, it is indicated that German NS specialists were involved in U.S. programs such as MKULTRA \cite{Koch04}. In the Soviet Union the question of continuing unconventional research was raised by the president of the USSR's Academy of Science in 1961, which was resolved positively \cite{Naumov93}. It can be assumed that the documents from Ahnenerbe archive were known to Soviet top leaders and to some extent they stimulated extensive studies in the '60s and '70s.

The research in this period was split into a few areas. First of all, the classic parapsychological experiments with a variety of phenomena exhibited by psychics were continued. Second, the program on the influence of EM radiation on biological objects received essential attention. This is the largest program with many sub-branches and funding, which eventually led to the emergence of the 'psychotronic weapons' discussed in media. Both the first and second programs had open and closed parts. And third, a new direction related to the 'high-penetrating emission of non-biological nature' was founded. In this review, we pay more attention to this research.

\begin{figure}[thp]
\centering
\includegraphics[width=0.38\textwidth]{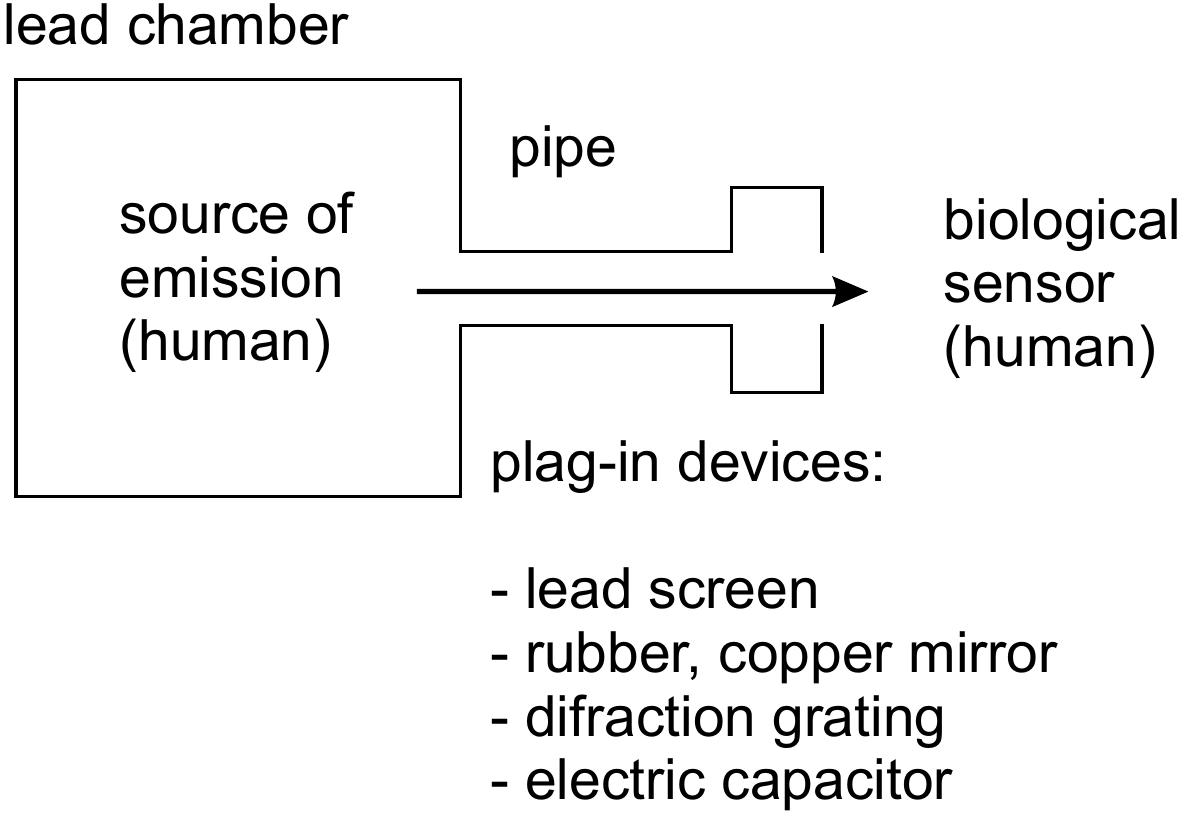}
\caption{\em \small Structure of experimental setup developed by S.Y.Turlygin. \label{fig:turlygin}}
\end{figure}

Let us return to the work of S.Y.Turlygin described in \cite{Vinokurov05}, \cite{Kajinski63} and partly in \cite{Tyrlygin42}. Turlygin was a well-known expert in the field of high frequency EM emission \cite{Turlygin51}, thus he possessed an excellent understanding of the subject of experiments. The structure of one of Turlygin's experiments is shown in Fig.~\ref{fig:turlygin}.

We will quote from \cite{Vinokurov05}:
\begin{quote}
'Among used removable devices, that supply a pipe, was a lead screen, which should presumably delay a radiation. This radiation should also meet a hard rubber or copper 'mirror' intended to reflect the radiation flux. By passing through the grating it must demonstrate a diffraction pattern -- the highs and lows of the energy density. In some cases, the alleged flow of radiation can pass between the plates of a capacitor. After going through a particular plug-in device, which transformed it, the radiation reached the test person. Thus, the study was intended to clarify a  purely physical picture of the phenomenon, and the test person has served as a bioindicator, the hypnotist -- as a  biogenerator of radiation. Analysis and summary of the results gave Turlygin a weighty reason to come to the conclusion that the lead shield holds the radiation. This was manifested by increasing time until the test person fell down in comparison with experiments in which the screen was not used. The experiments with mirrors confirmed the presence of radiation and 'optical' law of its reflection. Experiments with a diffraction grating allowed to determine the wavelength of the radiation - it was in the range of 1.8-2.1 mm. However, the electric field of the capacitor does not deviate the radiation.

Some concluding Turlygin's remarks are interesting. He wrote: 'From the viewpoint of physics, the most significant fact is that the behavior of a test person -- the duration of the exposition -- gives a clear optical picture, which can be explained only by the presence of radiant energy'. He continues, 'These experiments do not leave us in doubt about presence of radiation emanating from the human body' .... According to him, he [Turlygin] came to the conclusion that some of the properties and parameters of recorded radiation differ from electromagnetic radiation, for example, it does not deviate in the electric field of a capacitor' \cite[p.72]{Vinokurov05}.
\end{quote}

This work is extremely important, as it establishes the possibility of physical manipulation with 'biological' radiation. Thus, Turlygin's works in the 30s are very close with the works of Hieronymus, also in the 30s. Hieronismus also found that 'specific radionic' radiation has some properties of electrical and optical radiation. However, as we said in the previous section, neither Hieronymus nor Turlygin knew about each other. After the war, in 1952, Turlygin carried out a series of telepathic experiments with D.G.Mirza (Д.Г.Мирза) and opened a laboratory for the study of parapsychology in 1955. After the death of Turlygin in 1958, D.G.Mirza became a head of the laboratory. Interestingly, the question about continuing the research in this laboratory was decided, not only by the Institute of Biophysics, but most likely splashed in the Academy of Science. However, in 1958, no specific decisions or activities followed from this side.

The next phase of research begins in the 60s. As indicated by \cite{Naumov93} in 1961, a special meeting at the President of the USSR's Academy of Sciences academician M.V.Keldysh (М.В.Келдыш) considered the question about the laboratory led by D.G.Mirza. Academicians E.L.Asatryan, A.I.Berg, Y.B.Kobzarev, A.D.Minz, I.E.Tamm, A.A.Kharkevich (Э.Л.Асратян, А.И.Берг, Ю.Б.Кобзарев, А.Д.Минц, И.Е.Тамм, А.А.Харкевич) attended this meeting. The question about the laboratory was resolved positively. Apparently 1961 represents the beginning of a new program, since after 1961 the books of Kazhinskiy \cite{Kajinski63} (1963) and Vasiliev \cite{Vasilev62}, \cite{Vasilev63} (1962, 1963-- new edition from 1959) appeared. They contained results from the 30s and provided the first look into Soviet pre-war research. In the USSR, since the parapsychology belonged to the category of 'bourgeois mysticism', the publication of these books has meant the state gave a 'green light' for starting works in this area.

Another hypothesis about 1960 -- as the start of a new program -- is expressed in \cite{Vinokurov93}. The Minister of Defence R.J.Malinoskiy (Р.Я.Малиновский) received documents in 1960 about U.S. attempts to use telepathy in military purposes. Although these publications appeared later as unserious, they triggered the USSR and in turn the USA to new military programs. Further development of this Soviet program, in particular in marine, is related to many names, e.g. G.A.Sergeev (Г.А.Сергеев) \cite{Sergeev02}. He worked at that time in the Popov Higher Naval Academy of Radio Electronics in the field of hydroacoustics. Surprisingly, we found a number of his patents, e.g. \cite{Sergeev66} from 1964, which dealt with the measurement of biological potentials, in particular EEG. This would confirm the hypothesis of starting military programs after 1961 (EEG are used to measure brain activities also in Perov's experiments with rabbits).

Thus, in 1961 the Soviet unconventional research received new impulses. A number of separate programs both in open and closed areas were started, performed sometime by the same organization and researchers.

\textbf{I. Parapsychological works.} Based on publications of L.L.Vasiliev \cite{Vasilev62}, \cite{Vasilev63} and G.F.Plekhanov (Г.Ф.Плеханов) \cite{Plexanov04}, it can be argued that the orientation of research on biophysical fields and information transfer remained the same. We can give some examples of studies in the 60s. In 1963, Vasiliev, together with the Bechtrev's Brain Institute conducted successful sessions of telepathic communication between Leningrad and Sevastopol. They repeated results of such experiments from the 30s -- this was one of the first works intended to replicate experimental data. In 1965-1967, the group led by V.P.Perov (В.П.Перов) held telepathic experiments on rabbits, which had implanted electrodes in the lateral nucleus of the anterior hypothalamus. Rabbits-inductors and Rabbits-percipient were removed from each other on the distance of 7 km. The behavioral responses of rabbit percipients in relation to rabbit inductors stimulated by 1.5-2.5 volt signal were investigated. A series of 36 experiments have been carried out, which included 535 cycles. As pointed out by V.P.Perov \cite{Perov84}:
\begin{quote}
'... the number of matches is well over a half of the total number of cycles. The probability of obtaining these results randomly, determined by the significance level, is very low, which gives reasons to accept the hypothesis of existence of connection between rabbits, distant from each other at a distance of 7 km'.
\end{quote}
As claimed in \cite{Vinokurov93}, these rabbits were used as a system of biological communication in marine experiments, in particular with submarines.

From the 60s different academic institutes started the study of such phenomena as the 'dermo-optical perception' of Rosa Kuleshova (Роза Кулешова), and later in the 70s -- the telekinesis of Nina Kulagina (Нинель Кулагина). In 1969, a studio of science-documentary films created the first movie about Nina Kulagina and her abilities. This film marked the beginning of film making about parapsychological research. In the U.S. report 'Defense Intelligence Agency document: Controlled Offensive Behaviour - USSR, July 1972' it is stated that in 1967 the USSR had more than 20 centers for the study of paranormal phenomena, with a budget of \$21 million.

One of the most important points in the early 70s is the special commission established by the order of P.N.Demichev (П.Н.Демичев) -- the Secretary of the Central Committee of the Communist Party. The commission was requested to investigate psychic phenomena and to provide a scientific conclusion to Soviet leaders. Between the meeting in the Academy of Sciences and this commission was about 10 years -- during that time a sufficient amount of experimental results were collected to enable the assessment by the commission. This again points to the intensity of work in the 60s. The Commission recognized the reality of psychic phenomena. The report was published in 1973 \cite{Leontiev73}, one year later it was translated into several languages, including English, German, French and Italian. V.P.Zinchenko (В.П.Зинченко), one of members of the commission, said about this report:
\begin{quote}
'The main thing: we were able to articulate and to defend a principal point. The phenomenon exists. The communication channel is unknown. The affecting channel is unknown. Fans can look for!' \cite{Perevozchikov89}.
\end{quote}

The commission's work is the basis for the assumption that the interest of the Central Committee of CPSU for biological radiation and paranormal human's abilities (that was in the focus of the commission) appeared only in the first half of the 70s. This coincides with the early psychotronic movement and first symposiums (in 1970 and in 1973) as well as with public information about U.S. programs. It is likely that both countries have used each other in the arguments in favor of such studies and the struggle for funding. For instance, the Soviet representative in the International Association for the Study of Psychotronics was Prof. G.A.Samoilov (Г.А.Самойлов) from the Ministry of Internal Affairs \cite{Vinokurov93}.

\textbf{II. Impact of electromagnetic fields on biological objects.} The idea about the electromagnetic nature of biological radiation was rejected already in the 20s and 30s. However, studies on the impact of EM radiation on biological objects were continued and enlarged. Following the work of Michailovskiy and others from the 30s, it was found that the EM field, with certain parameters, can cause a variety of bio-physical and mental effects. It can be assumed that the psycho-physiological effects of microwave emission were actively investigated during the NS regime in Germany \cite{Greg12}, and after 1945 the technology was adopted by the countries-winners. According to another version, e.g. \cite{Tigran90}, the first mentioning of the fact that the pulse-modulated EM radiation can cause auditory hallucination was in 1956. Anyway, already in the 50s, the USSR and the USA had their own programs on studying the impact of EM fields on biological objects.

In 1953 the USA started the program 'MKULTRA' -- the CIA program, in which, according to the U.S. Supreme Court \cite{Court85}, about 80 institutions, including 44 universities, 12 hospitals, three prisons and 185 private researchers participated \cite{Horrock77}. As mentioned in the public documents, the program to some extent was motivated by the corresponding NKVD's program, with similar strategies of using psychotropic (e.g. drugs) substances and technical equipment.
\begin{figure}[thp]
\centering
\includegraphics[width=0.38\textwidth]{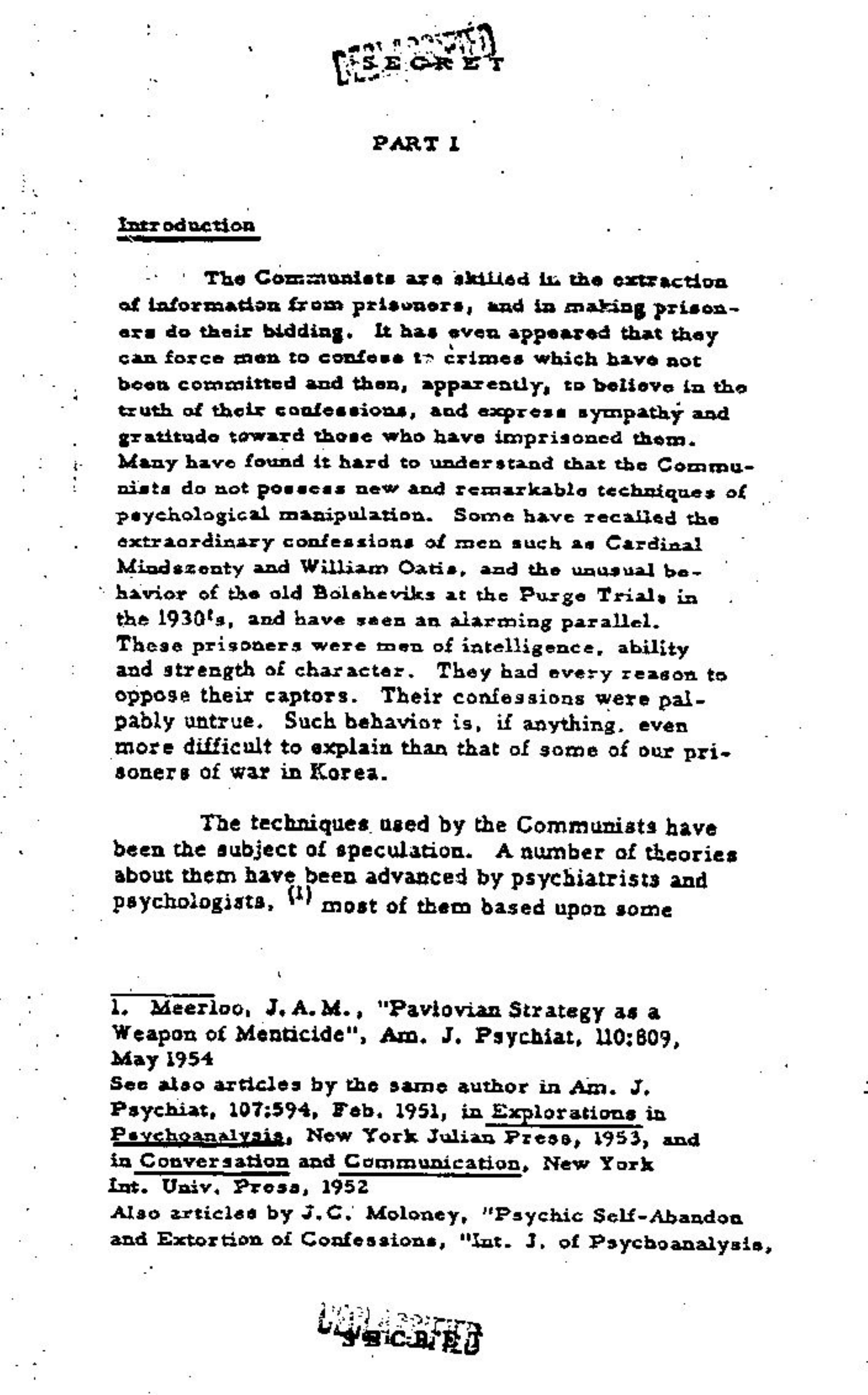}
\caption{\em \small Note on a possible connection between methods to influence the 'objects' from the project MKULTRA (1953) with Soviet works from the 30s, image from wikipedia. \label{fig:MKULTRA}}
\end{figure}

According to some authors, several German physicians with experience from NS concentration camps, were involved into MKULTRA \cite{Koch04}. This supports the hypothesis about works by Hitler's Germany in the field of psychotronics and the role of Ahnenerbe documents in Soviet programs. In the mid-70s, the program drew public attention; in 2011 some of these documents were declassified. Part of the MKULTRA program was also devoted to the influence of EM fields on the human physiological and psychological conditions \cite{Vorobievskij99}.

In the 60s and the 70s the Soviet Union had a large number of studies and research results on this topic. Y.A.Cholodov (Ю.А.Холодов) \cite{Holodov82} in 1982 so described the situation:
\begin{quote}
'Proceedings of the listed symposiums and conferences [more than 20 in 70s] constitute only a small part of the literature that are scattered in various journals and publications. After the publication of the collection of works on [subject] 'The effect of magnetic fields [MF] on biological objects' [1971], published by the Scientific Council on complex problems 'Cybernetics', Academy of Sciences of the USSR; several books and review articles were published. Dozens of dissertations were defended on specific issues of biological activity of magnetic field. There were published a number of bibliographies. In 1978 the second collection of the Scientific Council on complex problem 'Cybernetics' was published, entitled as 'reactions of biological systems to magnetic fields'. Today, there are more than four thousand references on biological effects of magnetic field, most of which appeared in the last decade. Approximately half of the publications devoted to the reactions of the nervous system to MF'.
\end{quote}
In the area of influencing EM fields on biological objects, several authors point to the application of research results in the form of new weapons (in the USA and the Soviet Union), see e.g. \cite{Kanduba95}:
\begin{quote}
'Over the past years, U.S. researchers have confirmed the possibility of affecting functions of the NS [nervous system] by weak EMFs [electromagnetic fields], as it was previously said by Soviet researchers. EMFs may cause acoustic hallucination ('radiosound') and reduce the sensitivity of humans and animals to some other stimuli, to change the activity of the brain (especially the hypothalamus and the cortex), to break the processes of formation processing and information storage in the brain. These nonspecific changes in the CNS [central nervous system] can serve as a basis for studying the possibilities of the direct influence of EMFs on specific functions of CNS' \cite{Holodov82}.
\end{quote}
Moreover, according to various authors \cite{Greg12}, \cite{Vorobievskij99}, there exist documents that would confirm the development of appropriate hardware in the USSR:
\begin{quote}
'The top-secret works were supervised by the twice Hero of the Soviet Union Marshal E.Y.Sawicki. It is said in one of the inquiries related to this invention, and stamped by the Institute of Radio Electronics of the USSR's Academy of Sciences: 'In 1973, the military unit 71592 of the city Novosibirsk, established the first installation 'Radioson' and conducted pre-tests. The positive results are reflected in the act of tests of this military unit ...'.... according to calculations made in 1974, the generator 'Radioson' can effectively 'treat' the city of about a hundred of square kilometers, plunging its inhabitants into a deep sleep -- and at a distance of up to 55 kilometers away from the transmitter' \cite{Greg12}.
\end{quote}
Other work \cite{Vorobievskij99} claimed that the act of testing, in addition to stamps of the military unit and the academic institution, was also signed by academician Y.B.Kobzarev and Dr. E.E.Godik (Э.Э.Годик).

The use of microwave radiation in the areas of SHF and EHF\footnote{Super High Frequency and Extremely High Frequency, 3--30 GHz and 30--300 GHz.} for impacting the psyche was covered many times in the press. The most well-known fact -- the discovery of 'strange antenna' in the office of President of the Russian Federation Boris Yeltsin:
\begin{quote}
'In the early 90's all headlines reported about this sensational discovery. In the media Yuri Malin confirmed: 'Experts have concluded that the antenna has been installed to provide psychological impact on the president'' \cite{Greg12}.
\end{quote}
Generally, the psychotronic SHF and EHF emission represents a topic for separate survey. On the basis of these studies, later, in the early 90s, some of the so-called non-lethal weapons have been developed \cite{DoDD96}. Although many of  the modern generators utilize the source of EHF waves \cite{Smirnov}, \cite{Smirnov10}, it was shown that 'high-penetrating' emission has some characteristic properties, which differ from electromagnetic radiation, see e.g. \cite{Kernbach13metrology}. For those who are interested in more detail, a good overview of research in the 70s and 80s in the area of biological radiation can be found in the book of P.Kneppo and L.Titomir \cite{Kneppo89}. It should also be noted that the impact of EM and other non-ionizing radiation on biological objects became a normal 'research topic', which at the moment doesn't belong to either parapsychology or to unconventional research. On contrary, this topic has become increasingly popular in today's scientific landscape, see e.g. \cite{Pilla1994}, \cite{4121246}, \cite{12929158}, \cite{Cherry2000}.

\textbf{III. Instrumental psychotronics.} Parapsychology of the 60s refers primarily to the field of psychology, for example, the members of Demichev' commission are well-known psychologists. However, technical aspects of biological radiation, in line with the Turlygin's experiments and books of Kazhinskiy and Vasiliev, were also very interesting for the technically-oriented scientific community. In 1965, the section of bio-information was organized by the Moscow's board of scientific and technical society of radio engineering, electronics, and telecommunications, chaired by I.M.Kogan (И.М.Коган). In 1968 the technical section of parapsychology and biointroscopy was organized under the central board of scientific and technical society of the instrument-making industry, led G.A.Sergeev.

It seems that the breakthrough in unconventional research and it's separation from parapsychology is related to the discovery of 'high-penetrating' radiation from non-biological origin. Even in the 30s it was already known that a 'high-penetrating' radiation possesses some properties of light and can be handled with prisms and gratings, as well as some properties of EM radiation and can be handled with a variety of screens and mirrors.

\begin{figure}[thp]
\centering
\subfigure[]{\includegraphics[width=0.28\textwidth]{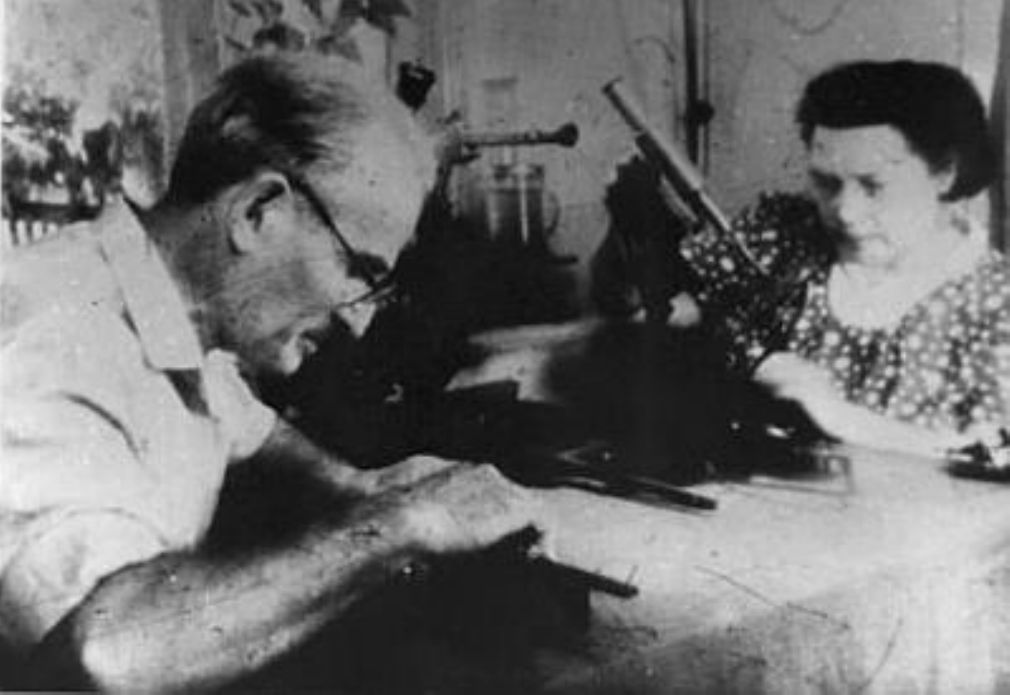}}~~~
\subfigure[]{\includegraphics[width=0.165\textwidth]{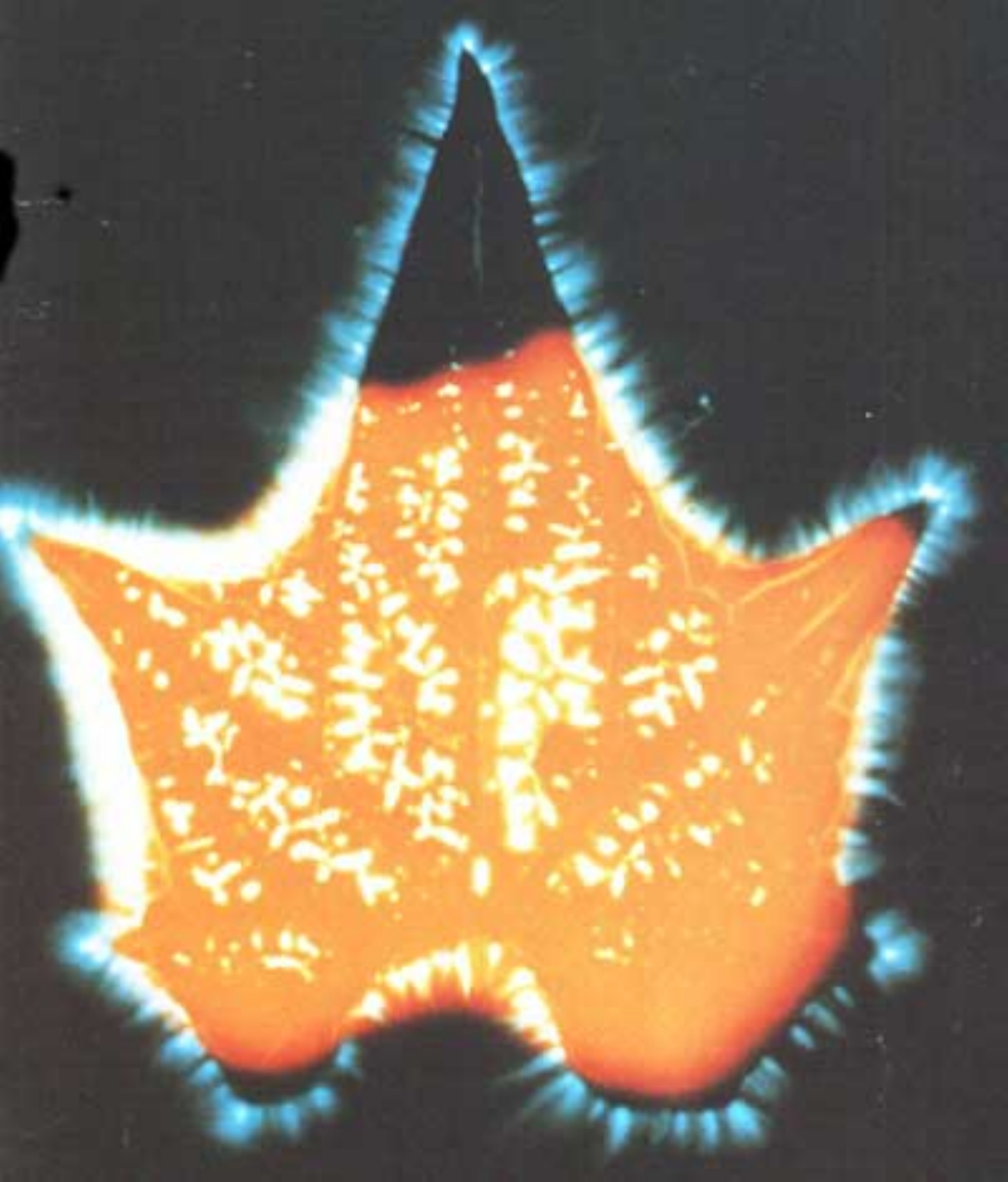}}
\caption{\em \small {\bf (а)} Family Kirlian, image from log-in.ru/articles/effekt-kirliana/; \textbf{(b)} Kirlian' image of a leaf with cut part, image from lebendige-ethik.net/4-kirlian-pribor.html. \label{fig:kirlian}}
\end{figure}

Before the 70s, Myshking's 'pondemotor forces' (related to spinning objects) and Chizhestvskiy's 'Z-rays' (from sun emission) were two approaches of generating non-biological 'high-penetrating' emission. However in that time, nobody connected these early works with the performed parapsychological research. The first approach of operator-independent detection of this emission is related to the Chizhevskiy-Velchover (Чижевского-Вельховера) effect, discovered in the 30s, and repeated in the 60s and later. This effect is about changes in metachromasia of corynebacterium several hours prior to changes of sun activities \cite{Habarova04}. The second approach was developed by the family Kirlian, they received a patent in 1949. In 1964 the book of Kirlians \cite{Kirlian64} appeared, which described the glow effect of objects in the strong EM field, see Fig.~\ref{fig:kirlian}. Interest in the Kirlian effect in the context of unconventional research emerged later, with the development of psychotronics, in particular, with the first conference on psychotronics \cite{Vilenskaj74}. The third approach of detecting the emission goes to N.A.Kozyrev (Н.А.Козырев) and is about a change of conductivity and mechanical properties in some materials and systems. The first 'unconventional' Kozyrev's work had been published in 1958 \cite{Kosyrev58}, however, since he was not in the framework of the state programs \cite{Shikhobalov02}, his work became known only after the death of the scientist in 1983. Thus in the early 70s, several approaches for generating and detecting the 'high-penetrating' emission without human operator were developed -- it was only a question of time before somebody started to connect these independent results. This process started in the mid-70s.

The first instrumental generators of non-biological 'high-penetrating' generators were created in 60s by Robert Pavlita in Czechoslovakia. Pavlita's talk at the First International Conference on psychotronics caused a furore, the conference itself was reported in central Soviet scientific journals of that period -- for example in the journal 'Science and Technology' for 1974 \cite{Sergeev74}. Here's what the magazine said:
\begin{quote}
'Experiments of Czechoslovak researchers R.Pavlita and D.Krmesski prove the possibility of remote impact to light moving objects. To enhance the impact, R. Pavlita offered a special device -- an 'accumulator' of energy. These 'accumulators' are made of different materials and have different shapes ... R. Pavlita found a number of other interesting properties of the investigated energy. Empirically, it is found that seeds of beans irradiated by this kind of energy germinated earlier than usual, the plant itself has evolved significantly faster than the control specimens ... R.Pavlita also discovered the accelerated deposition of aqueous suspensions under the influence of bio-energy. For example, if a water contaminated with industrial waste was filled into a container with metal shavings, irradiated by such energy, then 12 hours later the water becomes crystal clear. Furthermore, the chemical analysis shows that this is achieved with a very high degree of water purification. If the same water was 'in contact' with non-irradiated metal shavings, then the effect was not observed ... It is also necessary to point out the findings of the Estonian physicist T.Neeme. He experimentally confirmed an accelerated deposition of colloidal solutions under the impact of human bioenergy'.
\end{quote}
As the term parapsychology 'embarrassed' Soviet researchers due to a relation to 'inexplicable and mystical phenomena' -- which was contrary to the spirit of materialistic science in the USSR -- the term 'psychotronics' fitted well in the scientific picture of that time and has remained in use. On the contrary, the word 'psychotronics' is not widely accepted by Western parapsychologic community.

Intelligence services (KGB and DIA) were also interested in the works of R.Pavlita. As published in \cite{Kondakov95} and reprinted in \cite{Sandina95}, in 1972 Czech Ministry of Internal Affairs asked the USSR to assist in the investigation of Pavlita's devices. Two representatives: from Academy of Science -- A.Kitajgorodskiy (А.Китайгородский) and from KGB -- Y.Azarov (Ю.Азаров) arrived in Czechoslovakia. Kitajgorodskiy was a well-known opponent of such phenomena, see e.g. \cite{Reniksa73}, it seems this choice was motivated by two-side policy of the USSR in this area. He later wrote a number of sceptical reports. However, the KGB representative became very interested in these devices. Here is a quote from the book of David Sutter 'Age of Delirium: The Decline and Fall of the Soviet Union' (quote from the Russian version of the book):
\begin{quote}
'They talked about parapsychology and problems of communication. In the end, one that was lower and active, said that they are interested in experiments with a plate because they want to get an answer to a very important question. They said they have information that the Czech parapsychologist named Pavlita developed an apparatus for creating a biological field without human presence, and added that this discovery is of great interest, and that they need to find Pavlita, unfortunately, two years ago he disappeared, and they heard nothing about him. - When he died -- talkative one said -- he was not buried in the cemetery. We have checked all the cemeteries in Czechoslovakia. Block and Hronopulo suddenly lost all desire to participate in the experiment, because there was only one organization able to check all the cemeteries in Czechoslovakia' \cite{Satter05}.
\end{quote}
\begin{figure}[th]
\centering
\subfigure[]{\includegraphics[width=0.4\textwidth]{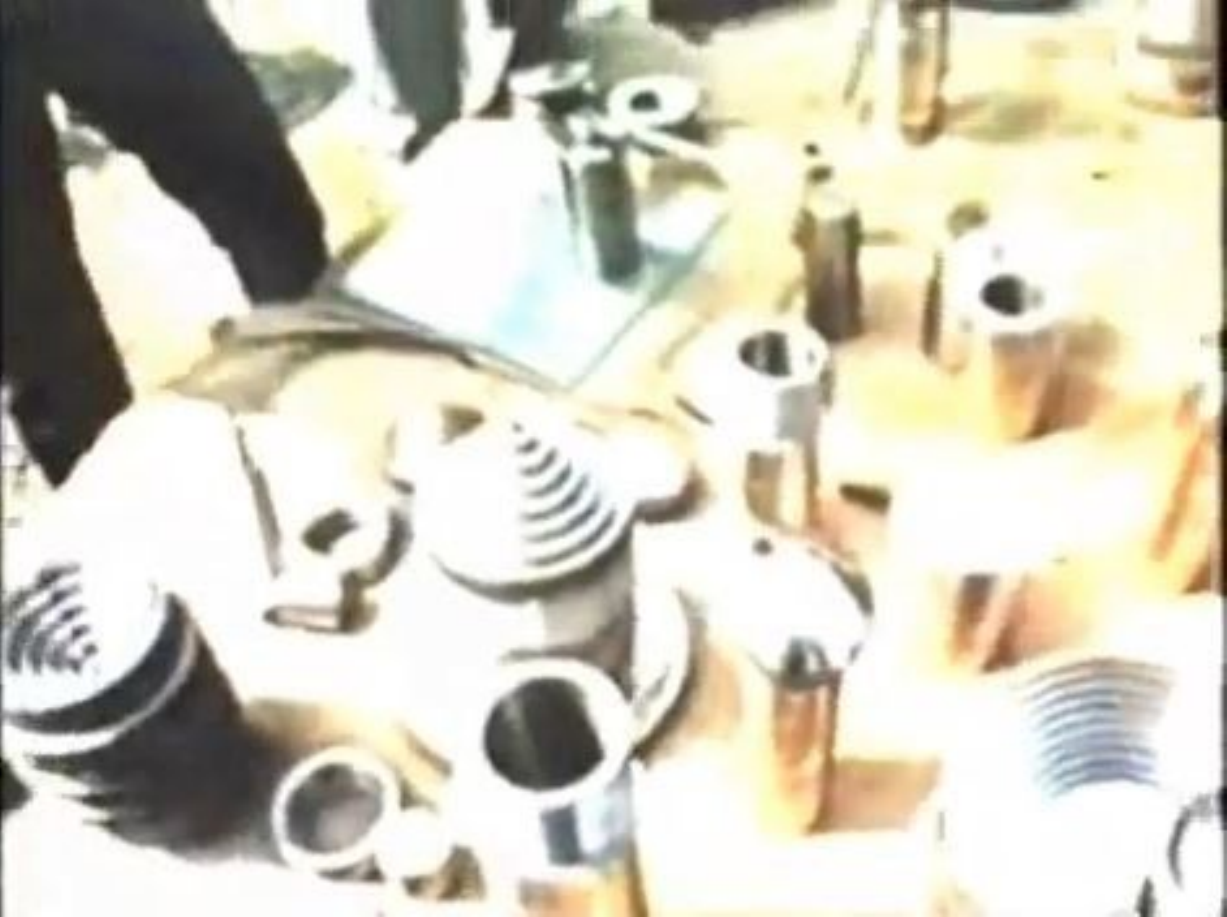}}
\subfigure[]{\includegraphics[width=0.245\textwidth]{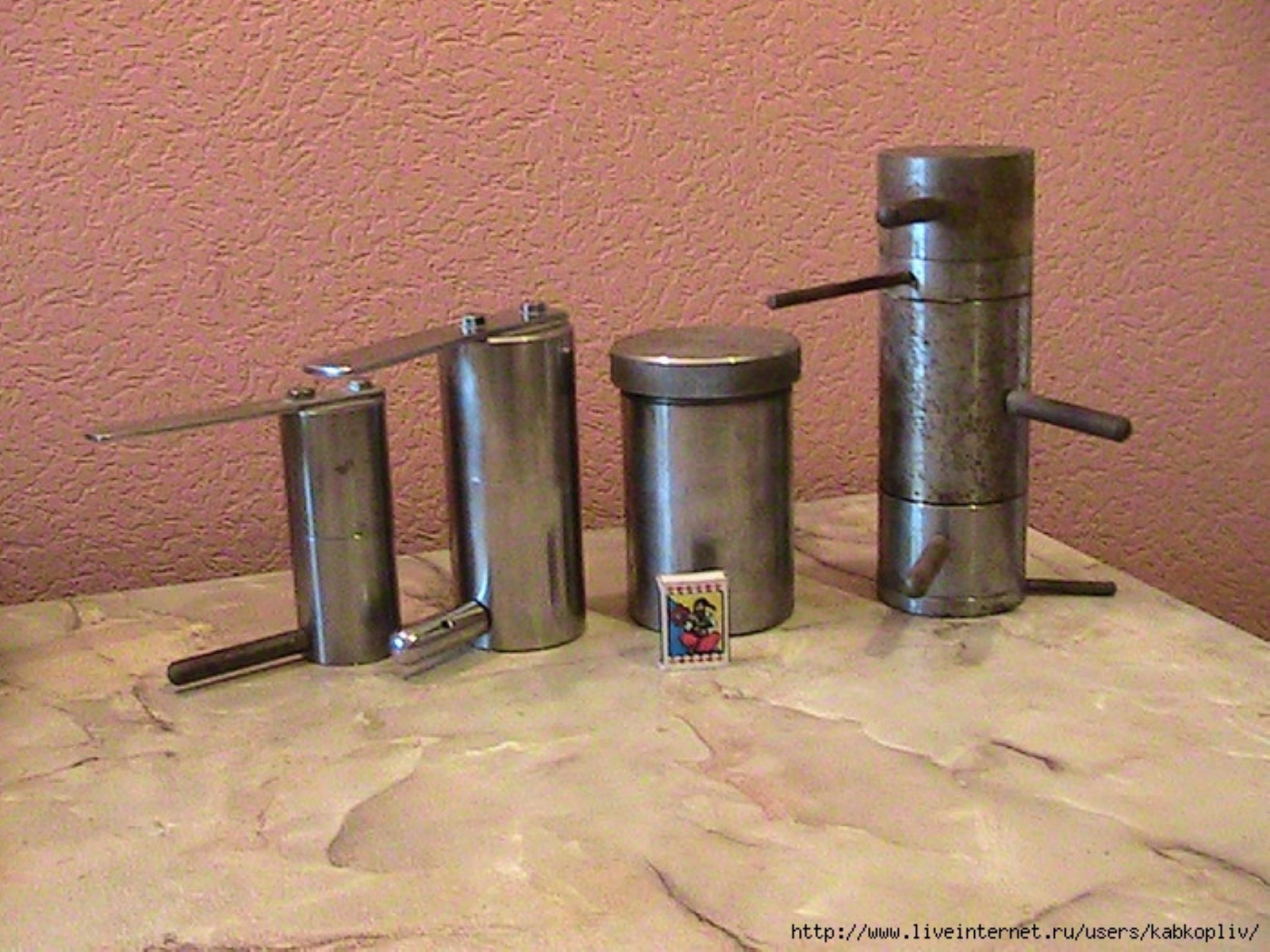}}~
\subfigure[]{\includegraphics[width=0.245\textwidth]{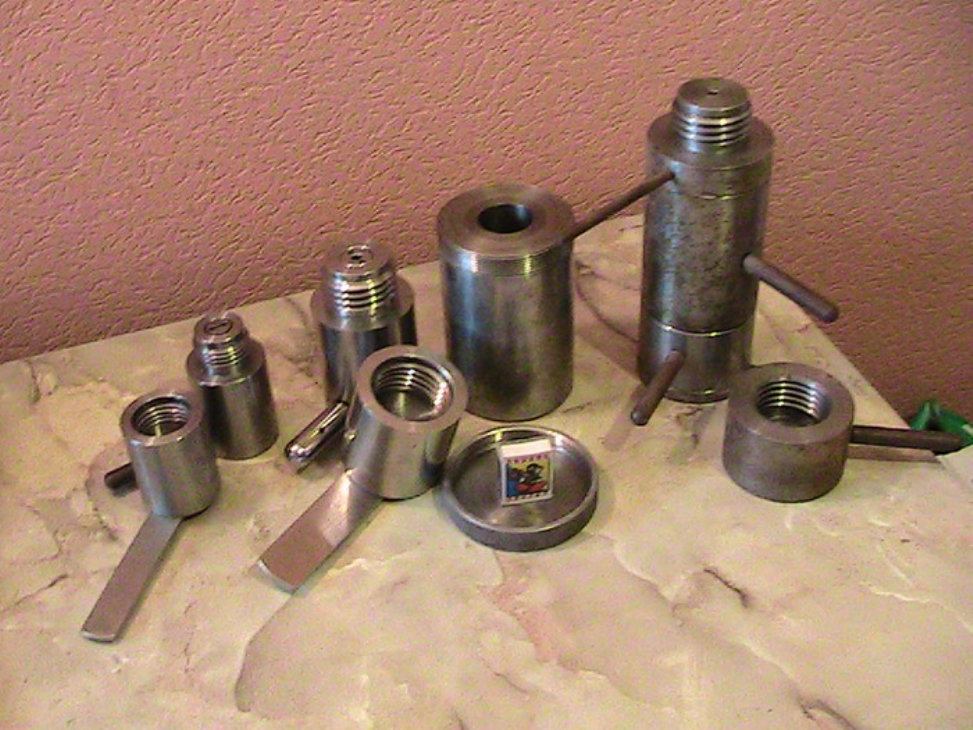}}
\caption{\em \small Different versions of the generator 'Cerpan', images are published with permission of kabkopLIV. \textbf{(a)} generators A.A.Beridze-Stakhovsky from the movie 'Lambada for Hiller'; \textbf{(b,c)} generators 'Cerpan', the comment from a modern manufacturer (www.liveinternet.ru/users/kabkopliv//post145559206/): 'The first three are original  samples produced for Beridze-Stakhovsky. The fourth one is the 'advanced' design created by us. The first two samples were made of stainless steel, the third one -- of titanium, the fourth one -- of steel. The main difference of the fourth sample -- it has complex internal structure associated with our ideas about the geometry of space and its energy properties'.
\label{fig:cyrpan}}
\end{figure}
As mentioned in \cite{PavlitaFoundation92}, the U.S. Defence Intelligence Agency (DIA) also indicated interest to Pavlita's work. He has registered many patents, however the secret of these devices he never opened (he died in 1991) -- so we can only assume the working principles of his generators. Based on some interviews \cite{Ostrander70}, it can be assumed the generators are passive devices that use so-called 'effect of shapes' ('effect of forms'):
\begin{quote}
'We saw a gallery of objects -- matt and shiny, rough and smooth, made of steel, bronze, copper, iron and gold, which were presented to us as 'psychotronic generators' ... Mystery generators -- claimed by Pavlita -- lies in their form. Important is also the material from which they are made. A position and shape of the material, which they are made from, are able to cause the desired effect. If the generator is designed properly, as claimed by Pavlita, it is able to accumulate bio-energy from all living things -- animals, plants, humans -- and then release it outside' \cite{Vinokurov93}.
\end{quote}
In the USSR, the first works of that time on the device-based generators were made by A.A.Beridze-Stakhovsky (А.А.Беридзе-Стаховский). His generator 'Cerpan', as shown in Fig.~\ref{fig:cyrpan}, is a passive device. The exact structure is unknown, both A.A.Beridze-Stakhovsky and R.Pavlita feared unethical use of their devices. According to some researchers \cite{Solodin03}, testing and verifying the work of these generators was controlled at the level of the Central Committee of CPSU.

\section{From 1980 to 2003}
\label{sec:19802003}

Works of the 80s and early 90s are primarily characterized by an appearance of centralized strategic programs at the level of the State Committee for Science and Technology at the Council of Ministers (SCST USSR). The role of the SCST USSR was to determine the main directions of the development of science and technology, planning and organization of the major developments of national importance for scientific and technical problems, the organization and introduction into production of discoveries, inventions, and the results of exploratory research (e.g. nuclear power). Thus, the strategic role of some unconventional areas, including instrumental psychotronics, was recognized. These works also received a significant amount of funding. Selection of 1980 (early 80's) as the last period in the development of unconventional research is motivated by the following reasons.

\begin{figure}[thp]
\centering
\includegraphics[width=0.5\textwidth]{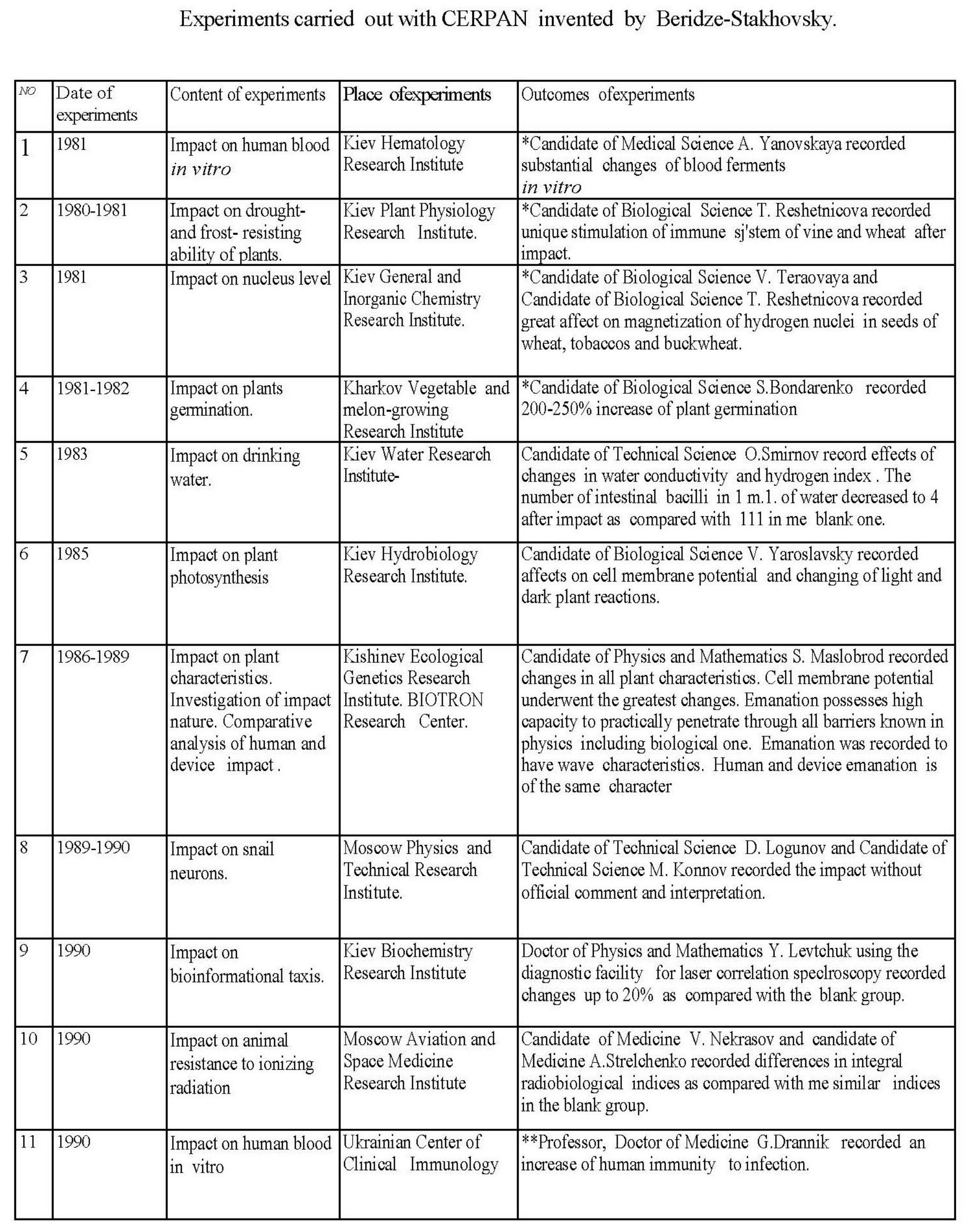}
\caption{\em \small Survey of experiments conducted with the generator 'Cerpan' from 1981 to 1990, data from www.liveinternet.ru/users/kabkoplivENG/, published with permission of kabkopLIV. \label{fig:cyrpanT}}
\end{figure}

First, the instrumental generators of the mid 70s, have been improved and the first experimental results of their work are obtained in 80s, as shown in Fig.~\ref{fig:cyrpanT} for generators of A.A.Beridze-Stakhovsky. Most of these works with this generator were made in Kiev. In general, Kiev, during this period, was one of the main centers of instrumental psychotronics.

V.A.Sokolova (В.А.Соколова) describes experiments that were carried out in 1984-1987, which are supervised by the USRR's Ministry of Health, the Ministry of Agriculture, and which took an interest in the Ministry of Defence.
\begin{quote}
'Works with torsion generators are mainly performed on the basis of Biophysical Laboratory of the Patrice Lumumba's University in the period from 1984 to 1987. In addition, our torsion generators have been tested in Moscow's leading institutions -- the Institute of Virology of Academy of Medical Sciences, Institute of Bioengineering, at the Gamal's Institute, Pharmacological Institute , NPO  'Volna', NPO of engineering, Institute of Crystallography, etc. Moreover, our team and generators are tested in a production environment in some farms of the Moscow area: cattle farm 'Kamenka' in Podolsk district, plant growing on state farms, and the farm 'Tarasovskay' in Pushkin area. With us at the farm 'Istra' worked ...'\cite{Sokolova2002}.
\end{quote}
As a part of these experiments various virological, biological, medical and agricultural works were carried out. The objects of these experiments were microorganisms, mice, plants, and even medical works with patients. According to Sokolova \cite{Sokolova2002}, already in 1986 more than 30 torsion generators were produced.

Second, due to contacts between authorities and psychics (it is assumed there existed contacts between Juna and Brezhnev \cite{Mularov99}), various research institutes received official orders to explore these phenomena. In 1980, the SCST and the Presidium of the USSR's Academy of Sciences commissioned the Institute of Radio Engineering and Electronics, AS USSR -- the lead agency of the Academy of Sciences to study weak signals -- the program of studying physical fields from biological objects in order to create fundamentally new methods of medical diagnostics. In particular, it included a study of physical fields from E.Y. Davitashvili (Juna) \cite{Gulaev84}. These works were carried out under supervision of U.V.Gulyaev and E.E.Godik \cite{Kozlova83}. Name of Y.V.Gulyaev is closely linked with other two names -- D.B.Kobzarev and N.S.Kulagina. The group of academics Gulyaev and Kobzarev investigated the phenomenon of Kulagina \cite{PhenomenD91}, which has been studied since 1977 in the St. Petersburg Institute of Fine Mechanics and Optics, where we meet the name of G.N.Dulnev (Г.Н.Дульнев) \cite{Dulnev92}. These works stimulated the development of a 'psychic line' of Soviet parapsychology.

Third, in the 70s and 80s United States had a number of programs in the fields of applied parapsychology and psychotronics. The first declassified (in 1995), and therefore the most well-known, state program of studying psi phenomena is the program on clairvoyance (non-local perception). Experts believe that there are actually a few independent programs in the CIA, counter-intelligence, naval and air forces. Two of them are the most famous. The first one was held on the CIA initiative since the early 1970s and since 1989 in SRI International (formerly known as Stanford Research Institute) and then from 1992 to 1994 at SAIC (Science Applications International Corporation), and later known as 'Anomalous Mental Phenomena'. The second program, known by various names -- 'Star Gate', was conducted under the auspices of counterintelligence. There are different opinions about these programs (although the CIA declassified 80,000 pages of text) \cite{Puthoff}, \cite{May96}. In Russia there are several books on this subject, such as the work of J.Mc Monigla 'Secrets of remote viewing' \cite{Mogil08}.

It is obvious that the USA and USSR 'competed' in this area -- research on the one side stimulated equivalent studies on the other side. Some representatives\footnote{ Interview of the head of Energy and Information Laboratory Russian NAST Academy major general FSO B.K.Ratnikov (Б.К.Ратников) to the magazine 'Security' on August 29, 2010.} of 'power structures' in the Soviet Union so characterized the 80s:
\begin{quote}
'In general, in 80s, in this country, it was created a system of well-organized and conspiratorial work to develop new methods and means of resolving interstate and internal political problems without involving intimidating power forces and damaging effects. It includes methods of obtaining timely information, other than the traditionally known'.
\end{quote}

\begin{figure}[thp]
\centering
\includegraphics[width=0.5\textwidth]{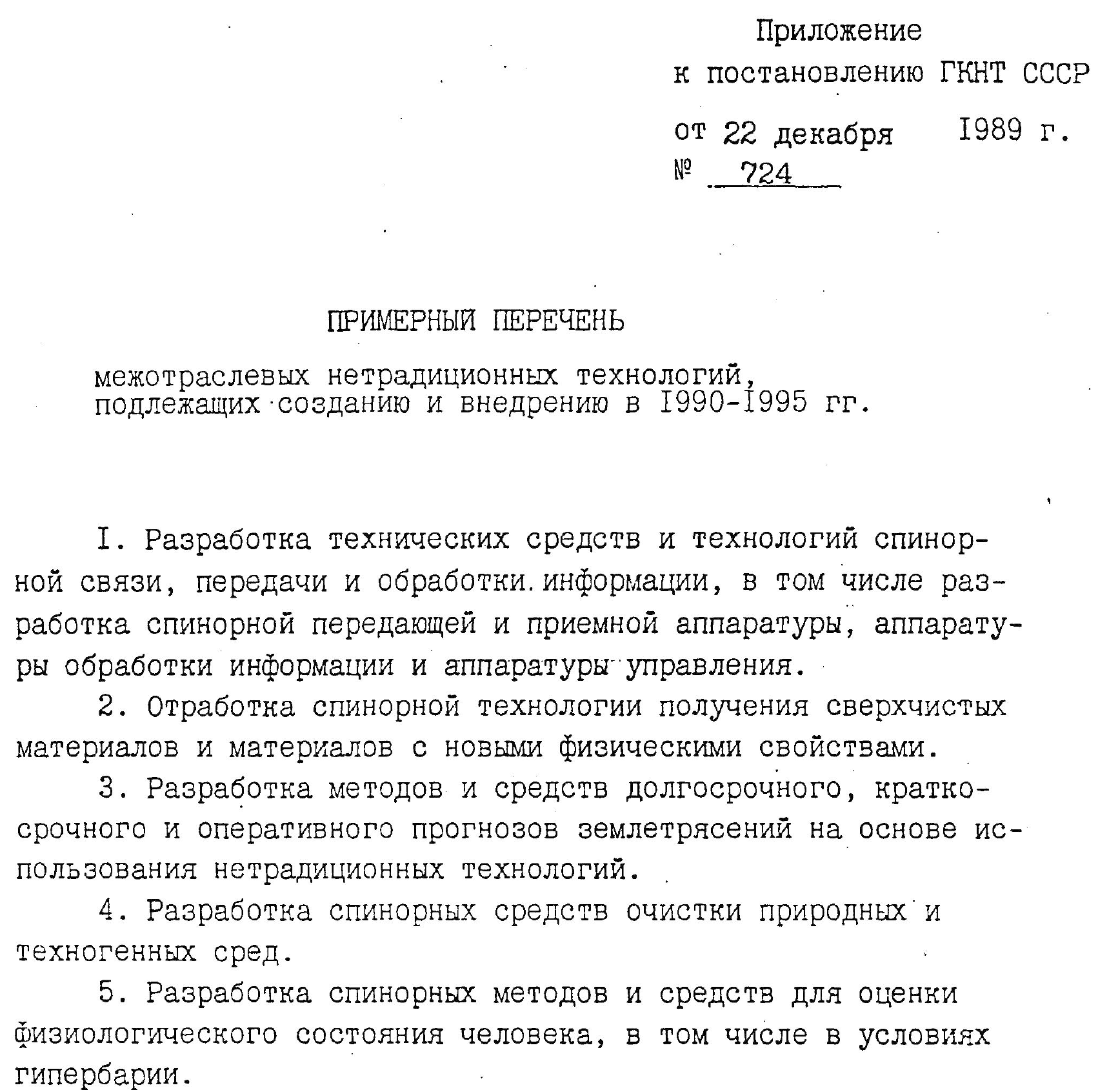}
\caption{\em \small Annex to the Resolution of the SCST USSR N724 from 22.12.89 on the establishment of a center of unconventional technologies at the State Committee for Science and Technology USSR, from \cite{Zhigalov09}.
\label{fig:centre}}
\end{figure}

\begin{figure}[th]
\centering
\includegraphics[width=0.5\textwidth]{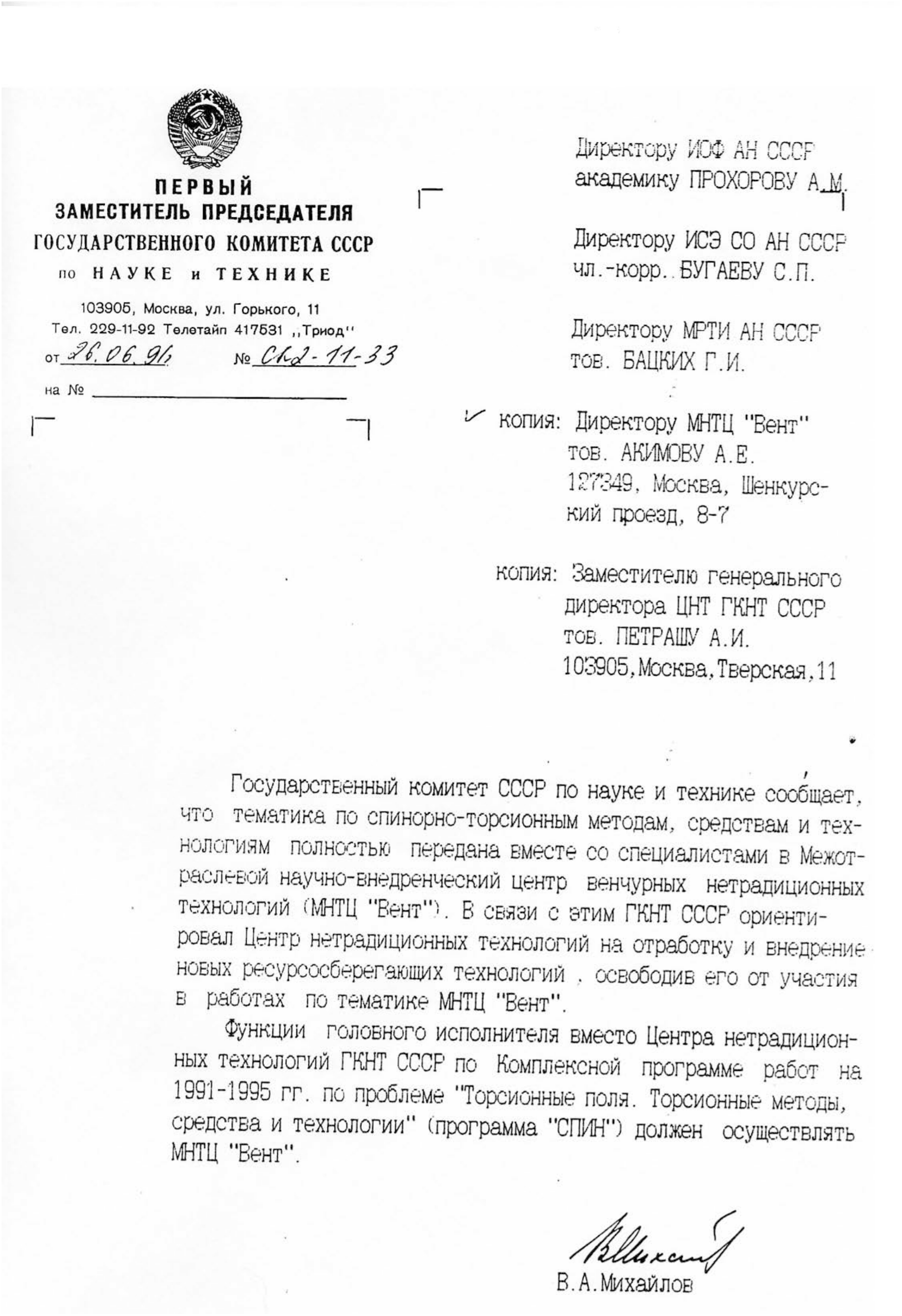}
\caption{\em \small Resolution of the SCST USSR  CK 2-11-33 establishing the ISTC 'Vent', from \cite{Zhigalov09}.
\label{fig:centre1}}
\end{figure}

Since the mid-80's the central coordinator of unconventional research became the State Committee for Science and Technology at the Council of Ministers (SCST USSR) with the direct participation of the Ministry of Defence and the KGB. In the middle of 1986, N.I.Ryzhkov\footnote{The last chairman of the Council of Ministers of the USSR -- the head of the government of the USSR.} (Н.И.Рыжков) on a memo about the perspectives of torsion technologies wrote a resolution: 'Take steps to organize the works' \cite{Akimov96}. Many authors\footnote{see psyterror.narod.ru, interwiki.info/index.php, www.uznai-pravdu.ru} point to the classified  document by the Central Committee of the CPSU and the USSR Council of Ministers N137-47 of 27 January 1986 about the program 'Management of living objects, including human'. For obvious reasons, the text of this resolution is not in the public domain, but this document has several indirect evidences. In May 1991, the Committee on Science and Technology at the Supreme Council of the USSR received a document from the member of the USSR' Academy of Sciences E.B.Aleksandrov (Е.Б.Александров), which states:
\begin{quote}
'From the mid-80's the defense agencies and KGB funded disjointed pseudoscientific closed developments, related to the problems of communication, weapons, and drug-free impact on human. In 1986, different groups were consolidated: they have entered into a document of CM [Council of Ministers]'\cite{Zhigalov09}.
\end{quote}
It is clear that Alexandrov is referring to the above mentioned decree of the Council of Ministers of the USSR from 1986. The instrumental psychotronics became a part of open programs on 22 December 1989. Decree N724 formed a Center of unconventional technologies that should deal exclusively with instrumental psychotronics, see Fig.~\ref{fig:centre}.
A.E.Akimov (А.Е.Акимов) became the head of the Center. From 1989 to 1991, these studies were open and coordinated by the Center. The unconventional works of the late 80s -- early 90s are mainly associated with this organization. In 1991, there appeared the well-known conflict between the Center and some representatives of the USSR's Academy of Sciences \cite{Zhigalov09}. On 26 June 1991 the SCST USSR created the interdisciplinary center of unconventional technologies 'Vent', which received all the functions, including financial ones, for coordinating the program, see Fig.~\ref{fig:centre1}. However, a week later, an 4 July 1991 the same SCST USSR issued the resolution about 'Vicious practice of financing ...', where it blamed the performed unconventional research and the Center. This is the oddest fact, where the same state organization within a very short period of time issued two distinctly different decrees, which had dramatic consequences for the whole country.

As it is repeatedly stressed in this work, the Russian unconventional research is characterized by a strong antagonistic position of supporters and opponents of this research, even in the same state agency. As stated by A.E.Akimov, see \cite{Smaga07}, \cite{Zhigalov09}, this program should be about 500 million rubles (more than \$500 million). It seems that several organizations claimed the coordinating role at the end of the 80s, including the Academy of Science, which traditionally dealt with these issues.

\begin{figure}[th]
\centering
\includegraphics[width=0.5\textwidth]{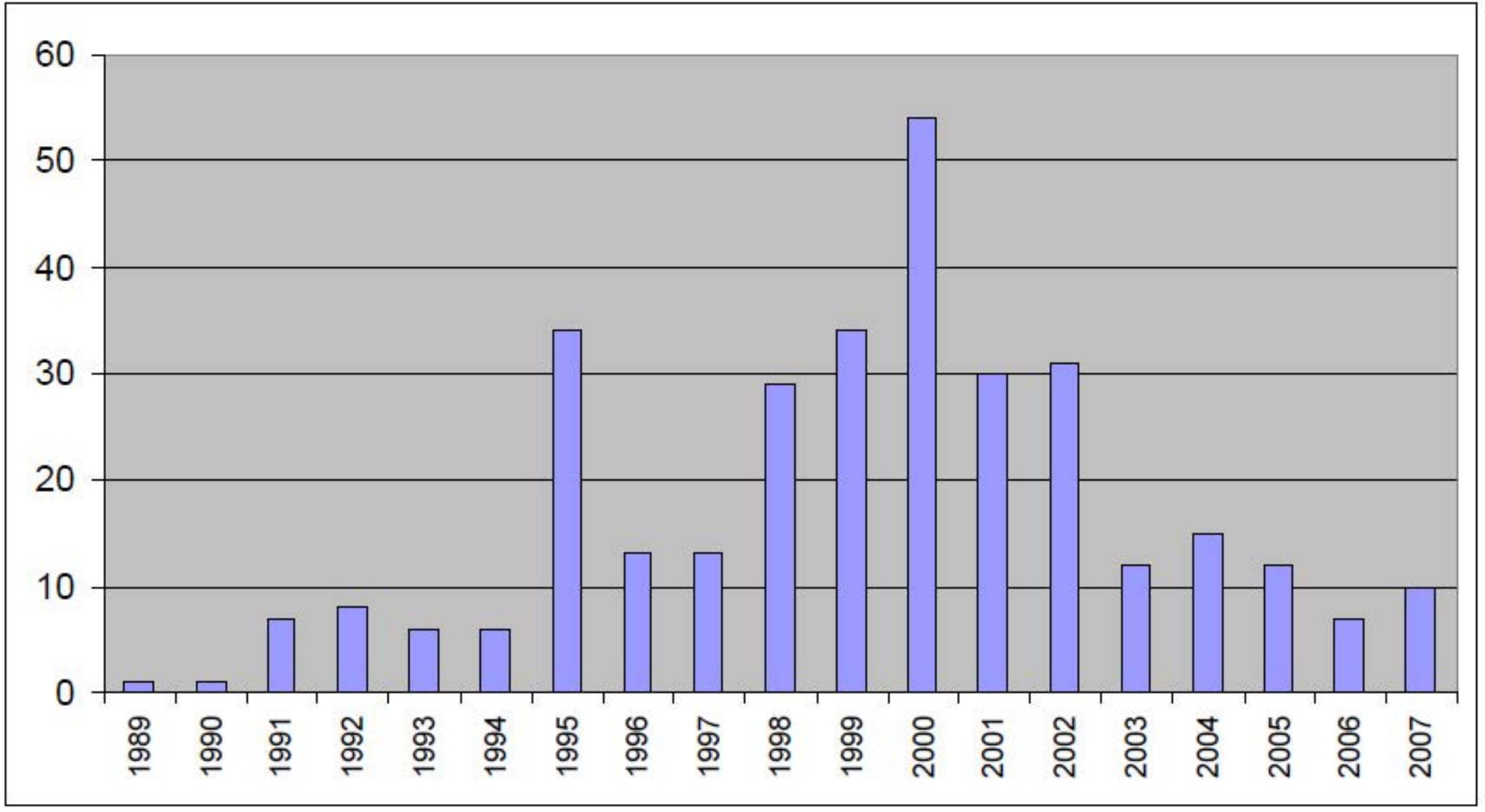}
\caption{\em \small Number of publications on the spin-torsion interactions, image from \cite{Zhigalov09} with permission of V.A.Zhigalov.
\label{fig:Zhigalov}}
\end{figure}

According to the statement of A.V.Bobrow (А.В.Бобров), published in \cite{Zhigalov09}, the center 'Vent' - in different forms -- existed until 2005, with funding and active works up to the end of the 90s. According to publications in \cite{Zhigalov09}, the work of the center has lasted at least until 1995, more recent publications are related to the 'International Institute of Theoretical and Applied Physics'. Publications, see Fig.~\ref{fig:Zhigalov}, show that the peak of research activity associated with instrumental psychotronics, continued until 2002-2003.

It should be noted that in addition to the program supervised by SCST USSR, there were several other activities relating to parapsychology. One of them is the Association of applied eniology created in 1989 by the Federation of Soviet engineers, led by F.R.Hantseverov (Ф.Р.Ханцеверов) \cite{Hanzeverov96}. As stated in \cite{Hanzeverov96}, there was a close cooperation with the International academy of energy science (MAEN). In 1991 a fund of parapsychology named L.L.Vasilev was created. From 1991 to 2000, the fund published the magazine 'Parapsychology and Psychophysics'. After 1991 a lot of organizations began to appear, such as the scientific committee 'Bioenergy', headed by academician V.P.Kaznacheev (В.П.Казначеев); the Center for psychotronics and traditional healing, led E.K.Naumov; the International committee of the socio-scientific human ecology and energoinformatics, led by Prof. V.N.Volchenko (В.Н.Волченко). In 1994, in St. Petersburg the Energy technology center, led G.N.Dulnev, was established.

According to many sources, 2003-2004 represents the end of this period in unconventional research. After 2000 the group of Kreml' psychics \cite{Kusina2005} was dissolved; the magazine 'Parapsychology and Psychophysics' was closed; many laboratories, see e.g. \cite{Smirnov10}, were closed in 2004; at the end of 2003 the military unit 10003, established in 1989 to explore the possibilities of military use of paranormal phenomena \cite{Ptichkin09}, was eliminated. Unfortunately, most of the organizations that arose during the period of the early 90's -- without proper funding and in terms of ideological pressure from the Russian Academy of Sciences -- no longer existed in the first decade of 21st century.

Summing up the section on the period from 1980 to 2003, it should be noted that with the collapse of the Soviet Union in 1991, these program are first reduced, and around 2002-2003 completely closed. Due to a conflict with the Academy of Sciences -- that is attributed by many authors to the financial conflict inside SCST USSR and not to scientific issues -- the whole unconventional research was announced as pseudoscience by the Academy of Sciences. Due to this conflict, even a publication of already obtained high-quality results was stopped -- researchers feared losing their academic jobs. It seems this conflict caused essential damages to Russian science in total. In this period we observe again a cyclical development with breaks in 1917, 1937 and 2003. Due to academic and non-academic researchers the instrumental psychotronics, denoted sometimes as torsionics, still continue to grow, but we cannot speak about government programs in Russia any longer.

\section{Devices and research of the period 1980-2003}
\label{sec:dev2003}

As shown in the previous section, the breakthrough in unconventional research was archived with the development of operator-independent ways to generate and to sense the emission, which was previously detected only by human operators. A critical amount of experimental data in the field of instrumental generators appeared first in the 80s. If in the 70s mostly passive generators, such as of A.A.Beride-Stakhovskiy, were developed, already in the mid-80s at least two active generators were present: one developed by A.F.Ohatrin (А.Ф.Охатрин) \cite{Vorobievskij99} (according to some claims, the Okhatrin's generator existed before 1982) and another one developed by A.A.Deev (А.А.Деев) \cite{Boldureva09}, \cite{Sokolova2002} (author in \cite{Boldureva09} claims that devices of A.A.Deev were tested at the Institute of Clinical and Experimental Medicine, Academy of Medical Sciences (Director V.P.Kaznacheev) in 1981). The Deev's generator was based on spin-polarized materials \cite{Boldureva09}.

In parallel with practical developments, researchers tried to understand the working principles of their devices, and often, not from theory to practice, but rather from practice to theory. Successful instruments were analyzed; researchers then tried to build a theory that would explain how they work. When in the experimental area a measurable progress was observed, in the theoretical area many questions remained unanswered. In the period from 1980 to 2003, a large number of such theories appeared; they often required reviewing of all the physics from the 16th century. A short overview of that research can be found in \cite{Kernbach13metrology} -- although a detailed analysis of all those approaches would require a much larger work. Here we would like to focus on three interesting topics that emerged during that period: 1) research on passive structures (the effect of shapes/forms); 2) the works, performed within the state program through the center 'Vent', which received a solid grounding in the theory of G.I.Shipov (Г.И.Шипов); 3) studying the impact of generated emission on a variety of materials and the development of instrumental sensors.

\begin{figure}[th]
\centering
\subfigure[]{\includegraphics[width=0.095\textwidth]{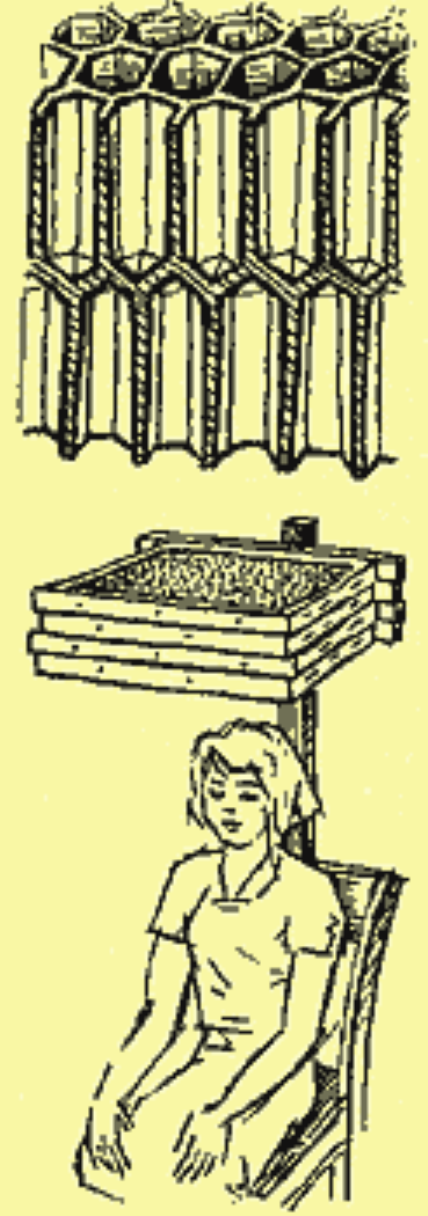}}~~~~~~
\subfigure[]{\includegraphics[width=0.2\textwidth]{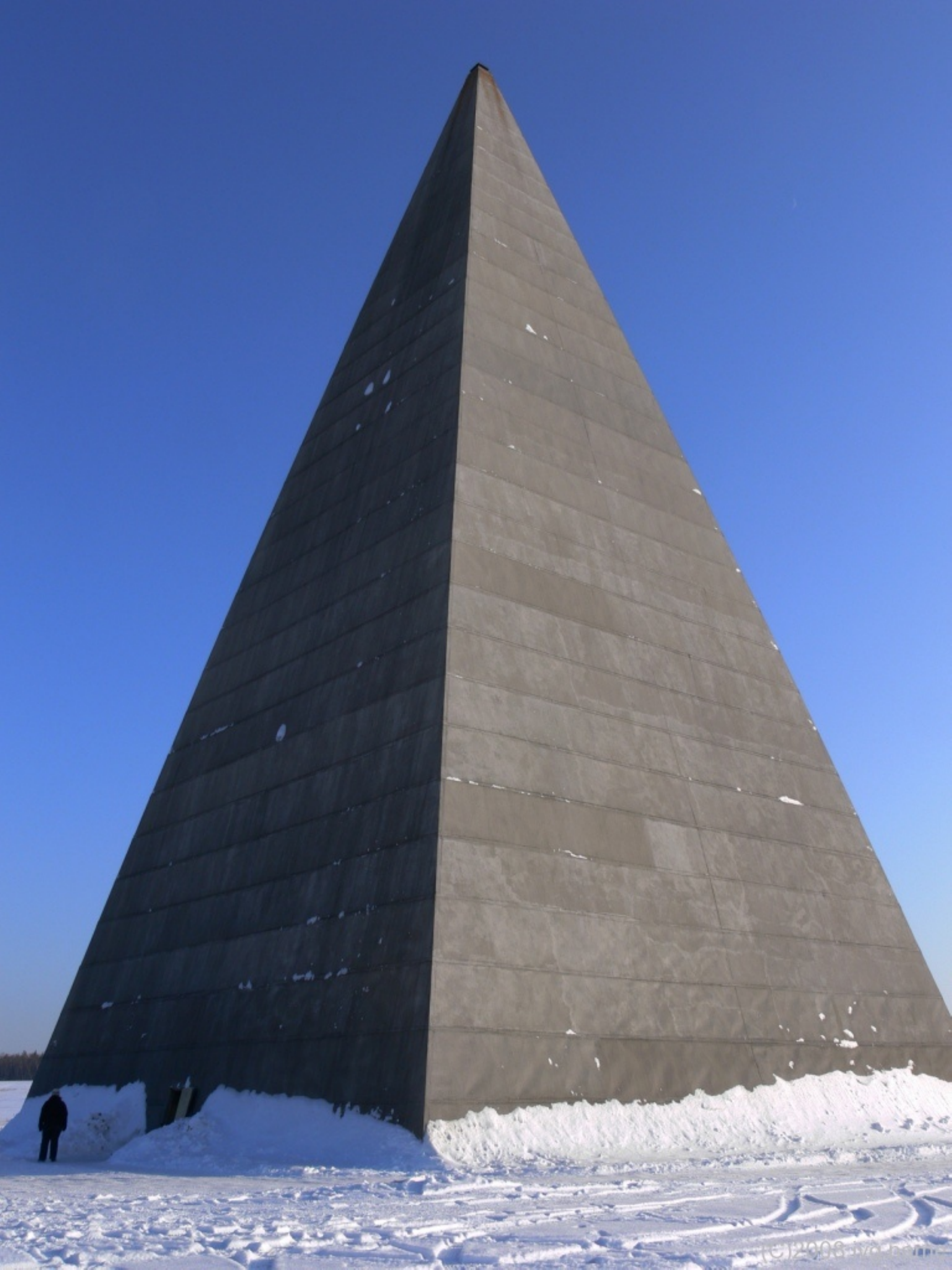}}
\subfigure[]{\includegraphics[width=0.24\textwidth]{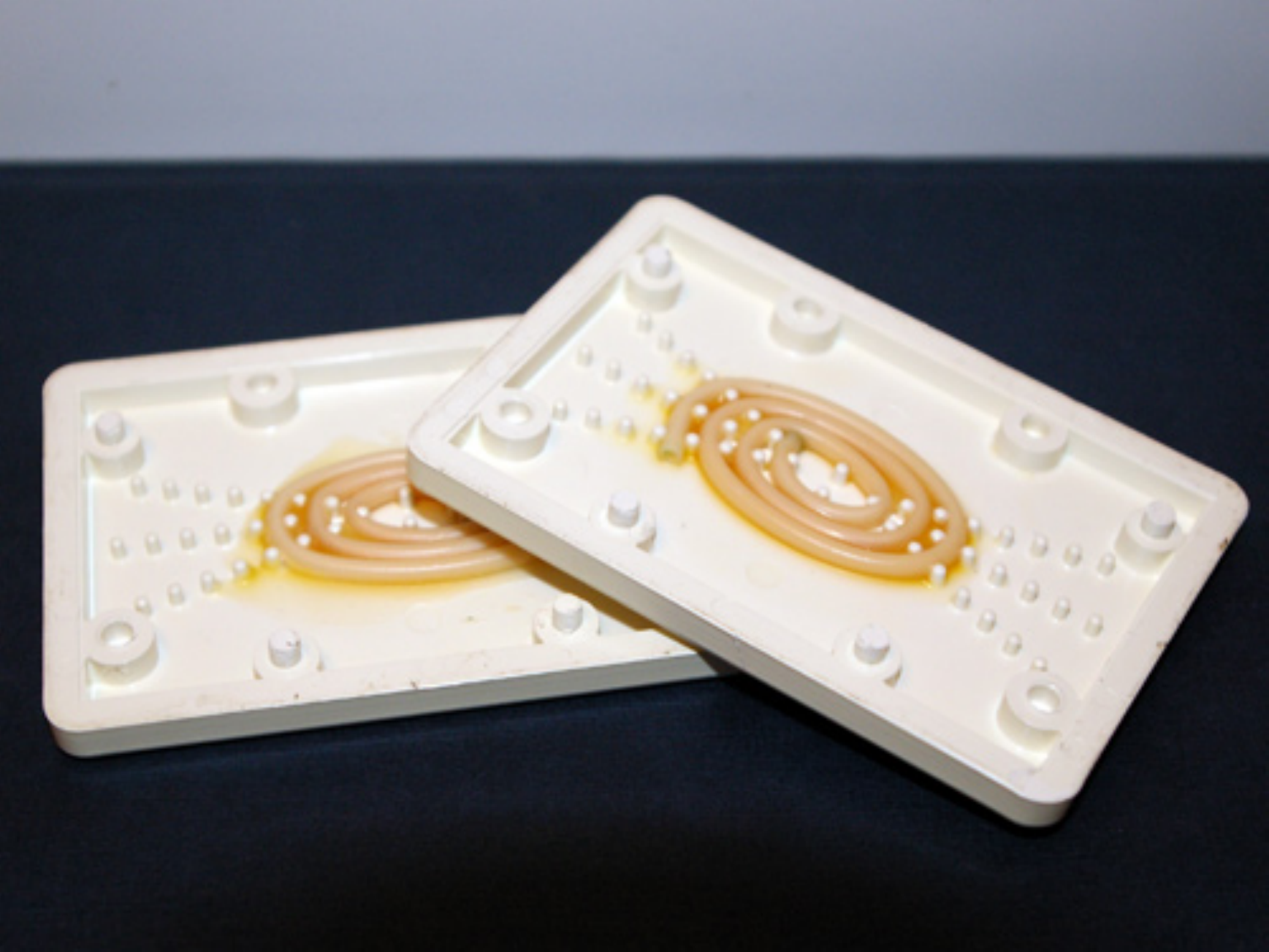}}~
\subfigure[]{\includegraphics[width=0.26\textwidth]{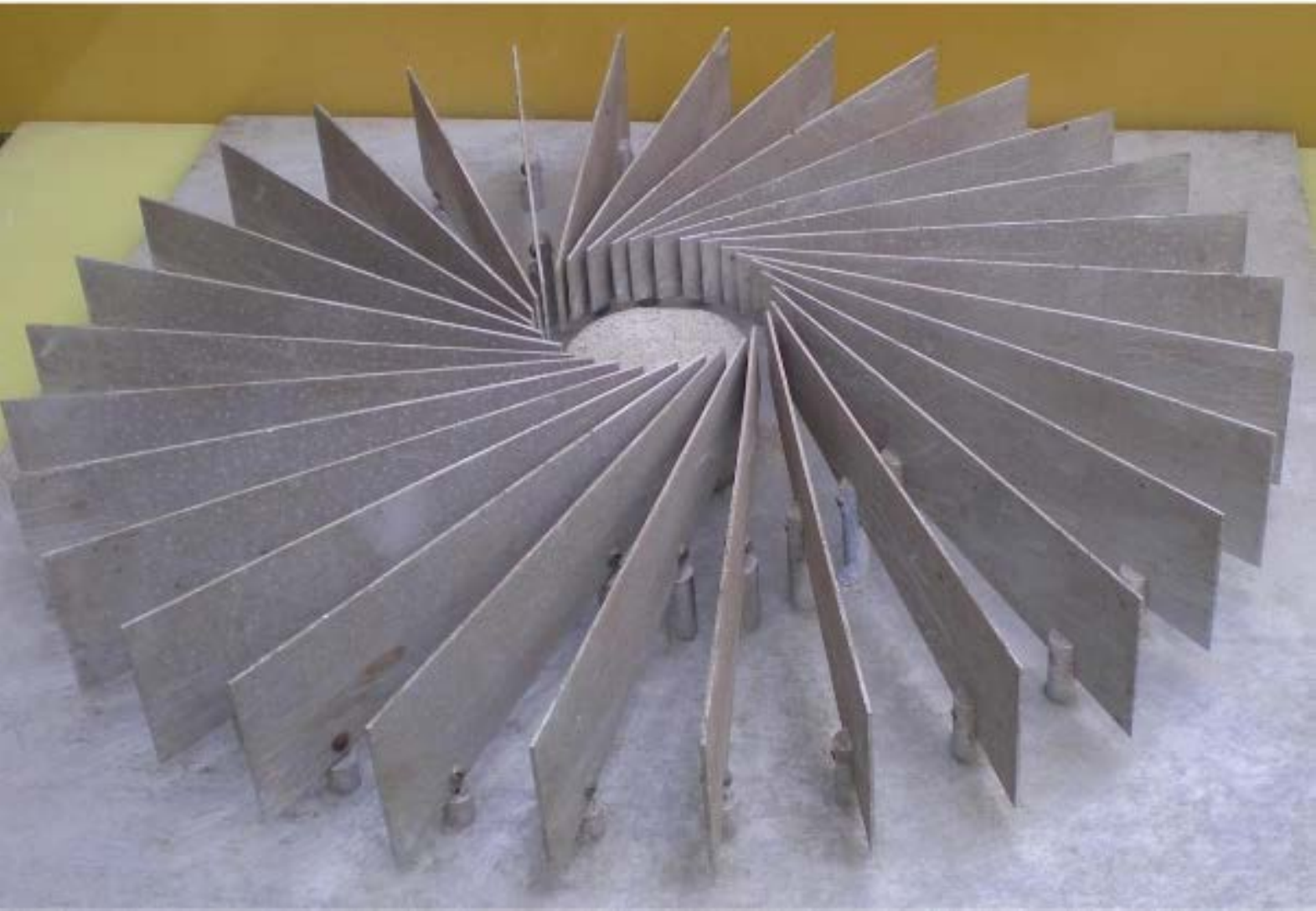}}
\subfigure[]{\includegraphics[width=0.4\textwidth]{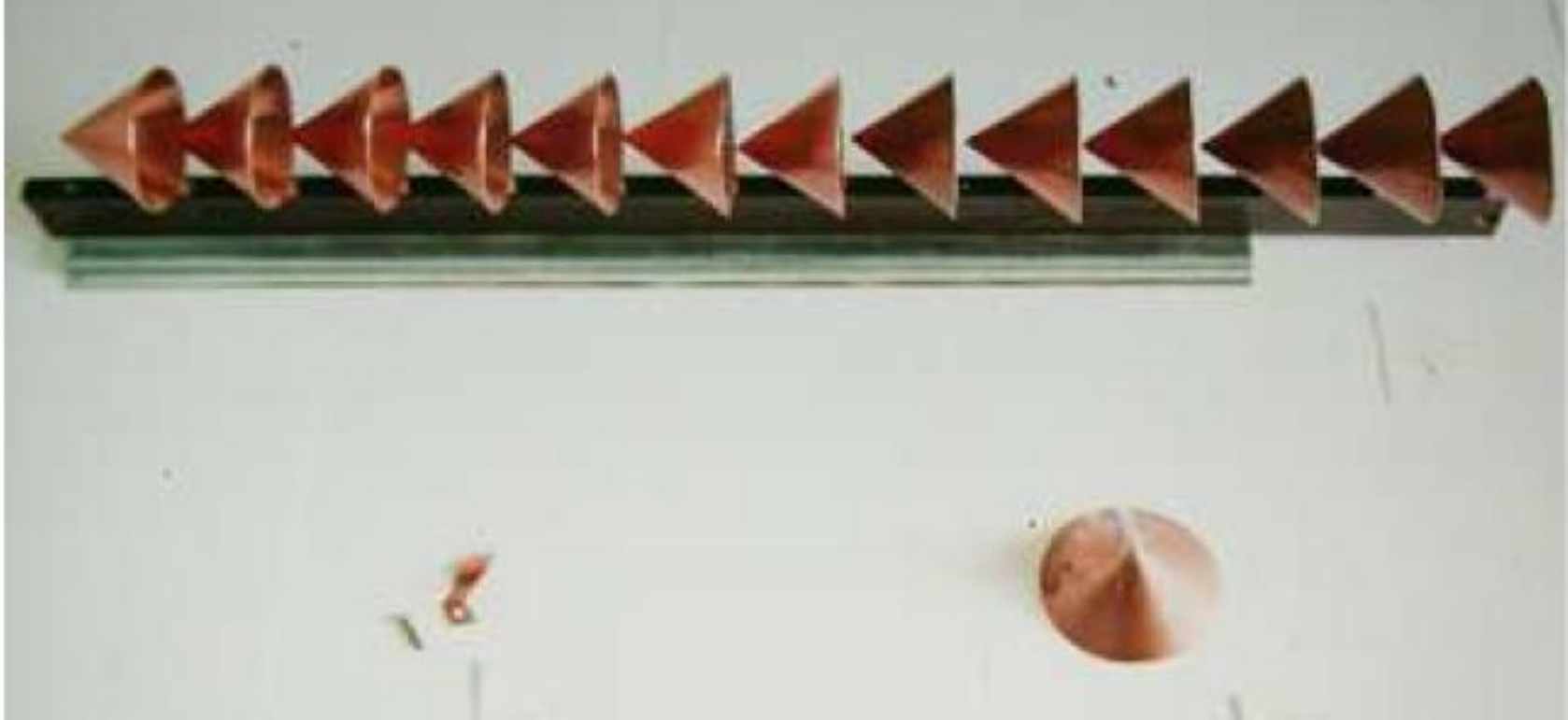}}
\caption{\em \small \textbf{(a)} The effect of hollow structures, image from \cite{Grebennikov98}; \textbf{(b)} Golod's Pyramid at Riga highway near Moscow, image from ivg.name/2009/02/02/pyramid/; \textbf{(c)} Passive generator 'Gamma-7' of A.F.Ohatrina, image from interwiki.info; \textbf{(d)} The 'Veinik hedgehog'; \textbf{(e)} Smirnov's passive generator, image from \cite{Zhigalov11}.
\label{fig:passive}}
\end{figure}

\textbf{Passive structure (shape/form effect)}. In 1983, at a meeting of the Novosibirsk's Branch of the Entomological Society of AS USSR V.S.Grebennikov (В.С.Гребенников) introduced the discovered 'effect of hollow structures' \cite{Grebennikov84}. The magazine 'Tekhnika Molodezhi' in N6 for 1984 wrote:
\begin{quote}
'Physiologists, physicists and doctors are deeply interested in this phenomenon. We are carrying out research to find instruments that could capture and detect this effect ... its nature (of the phenomenon) is not clear, even approximately' \cite{Grebennikov84}.
\end{quote}

Although the effect of hollow structures was known prior to this publication, e.g. from the French authors of the 60s and 70s \cite{Belizal65}, \cite{Pagot78}, both the paper and works of V.S.Grebennikov \cite{Grebennikov98} have stimulated a lot of research in the 80s and in the 90s: 'Veinik hedgehogs' \cite{Veinik91}, 'Cylinders' \cite{Kovtun02}, research on effects of pyramids \cite{Cherednichenko99}, \cite{Makin02}, \cite{Uvarov05}. The first reference and research on 'Veinik hedgehog' refers to the period before 1981 \cite{Veinik81}, but this manuscript, apparently, had not been published in the '80s. Widely known are Golod's pyramids \cite{Leskov00}, which from the beginning of the 90s were built in many cities of Russia, which triggered a broad public debate.

\begin{figure}[th]
\centering
\subfigure[]{\includegraphics[width=0.295\textwidth]{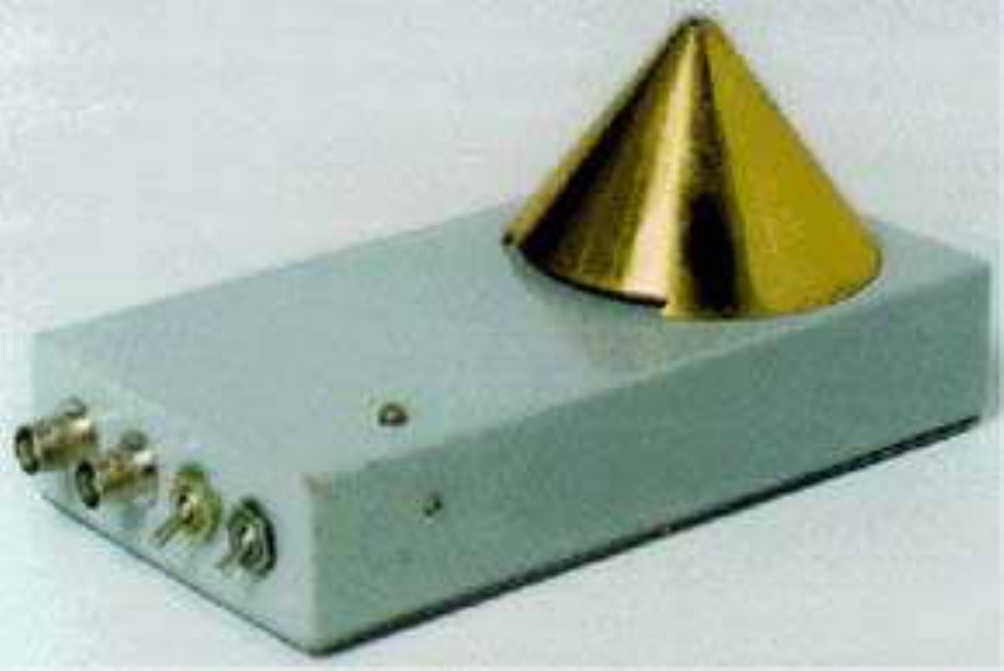}}~
\subfigure[]{\includegraphics[width=0.19\textwidth]{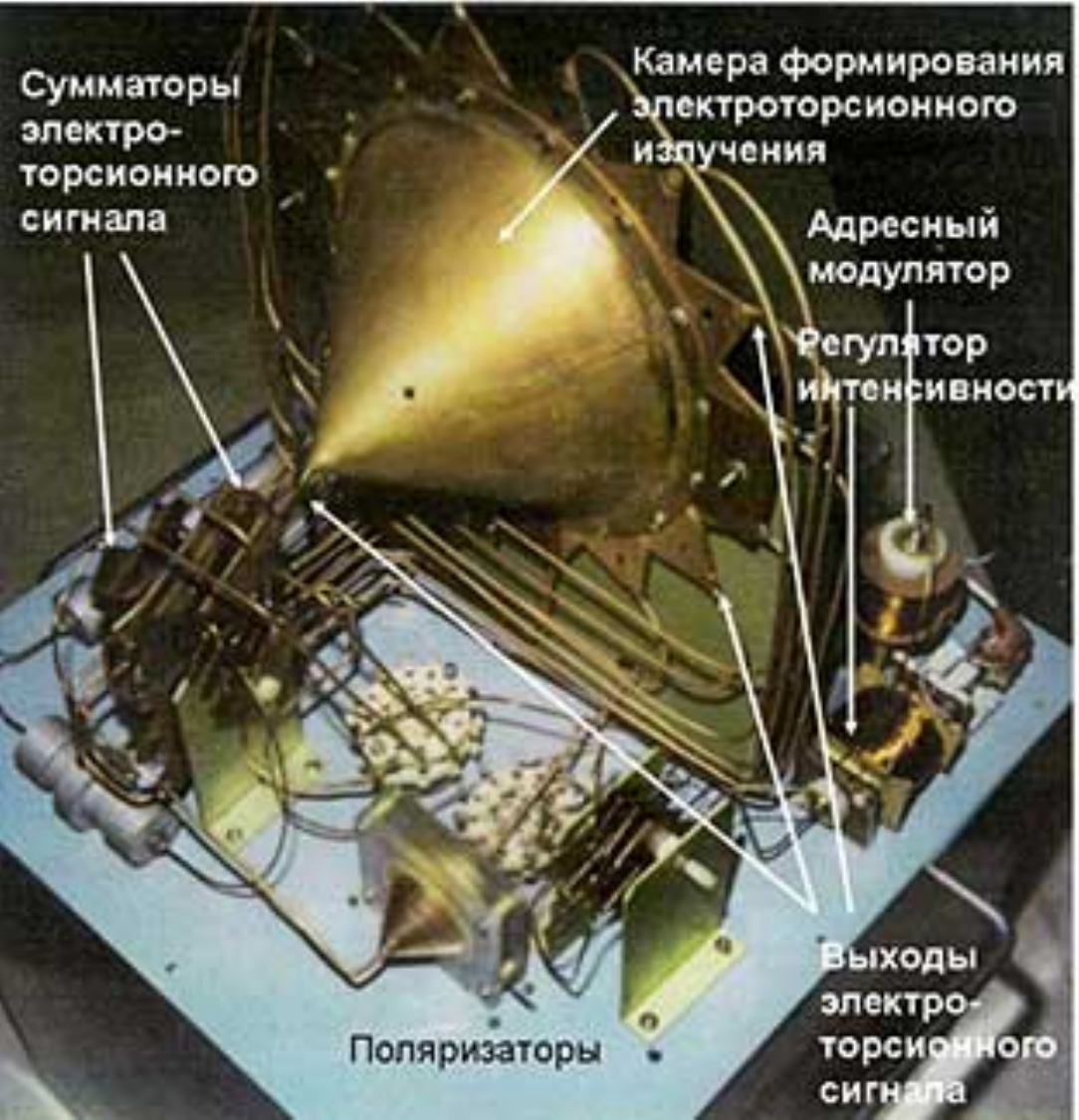}}
\subfigure[]{\includegraphics[width=0.24\textwidth]{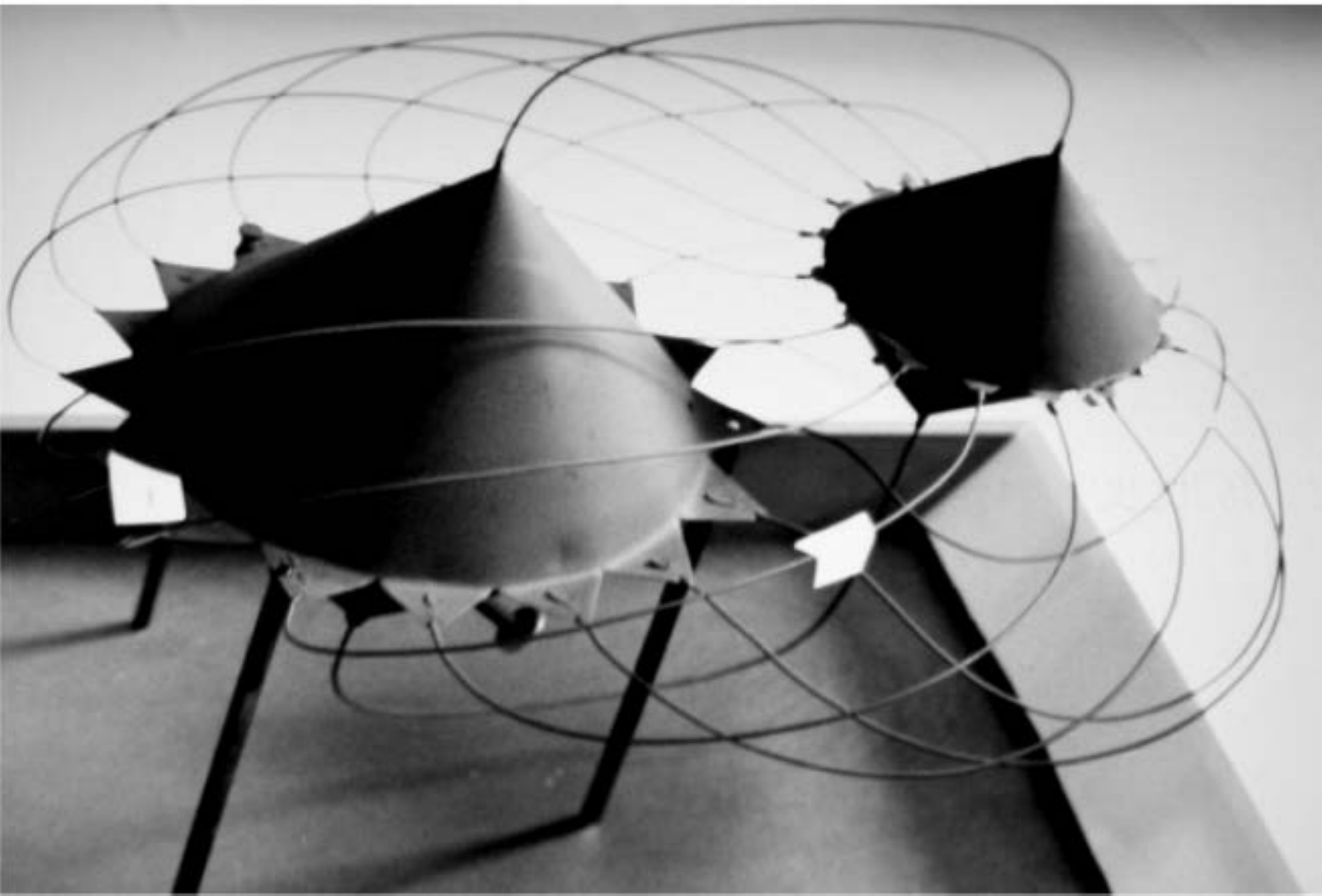}}~
\subfigure[]{\includegraphics[width=0.24\textwidth]{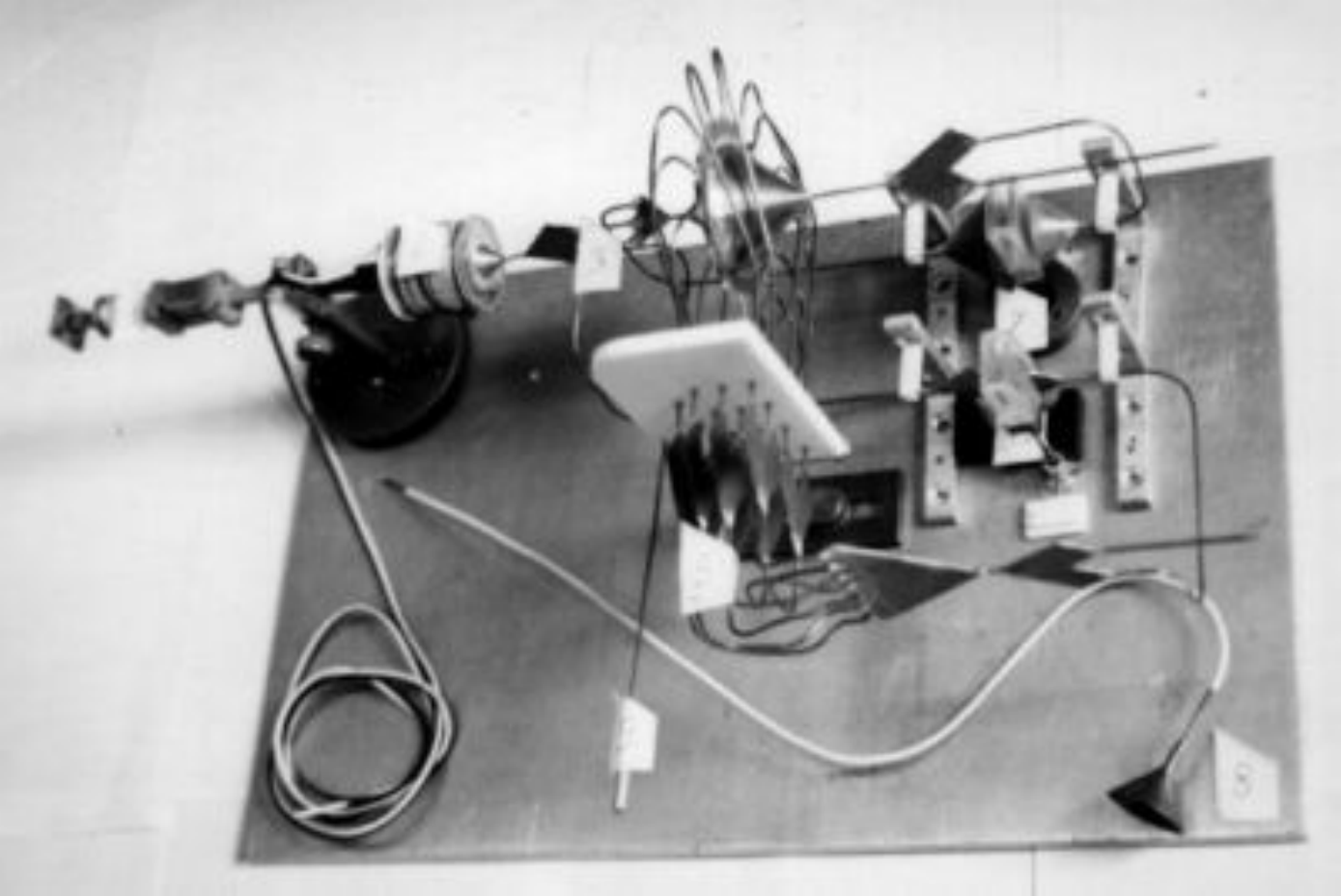}}
\caption{\em \small Examples of instrumental generators of a high-penetrating radiation developed at NII MP or under contracts with the center 'Vent': \textbf{(a,b)} small and large Akimov' generators, \textbf{(c,d)} generators of A.Y.Smirnov, images from \cite{Zhigalov09}.
\label{fig:aktGen}}
\end{figure}

It is assumed that passive structures -- based on so-called shape/form effect -- are similar to active instrumental generators. First passive generators were in fact objects of complex geometry. There is even such a notion as 'Pavlita geometry' \cite{Donovan13}, which should represent the basis of passive generators. Active generators, as shown in Fig.~\ref{fig:aktGen}, use the effects of forms. Some of such famous structures are shown in Fig.~\ref{fig:passive}.

\textbf{Works within the center 'Vent'.} Since the center 'Vent' was planned as a central coordinator of the program, all real works were performed by subcontractors with different research organizations. According to A.E.Akimov \cite{Smaga07} about 20 such organizations were involved after the collapse of the USSR in 1991. After the funding of programs was reduced, the core of researchers remained connected within the network. They organized and performed special conferences and created various research groups on the territory of CIS countries. Therefore, the works within the center 'Vent' mean, in fact, a very broad section of topics, of different authors, and of different organizations. The majority of developments in the instrumental psychotronics in Russia from 1991 to 2013 was united in this movement. The 'Vent' research program was related to five following topics, see Fig.~\ref{fig:centre}: communication, material research, new sensing technologies (e.g. for forecasting earthquakes), water cleaning, medical diagnostics.

\begin{figure}[th]
\centering
\includegraphics[width=0.45\textwidth]{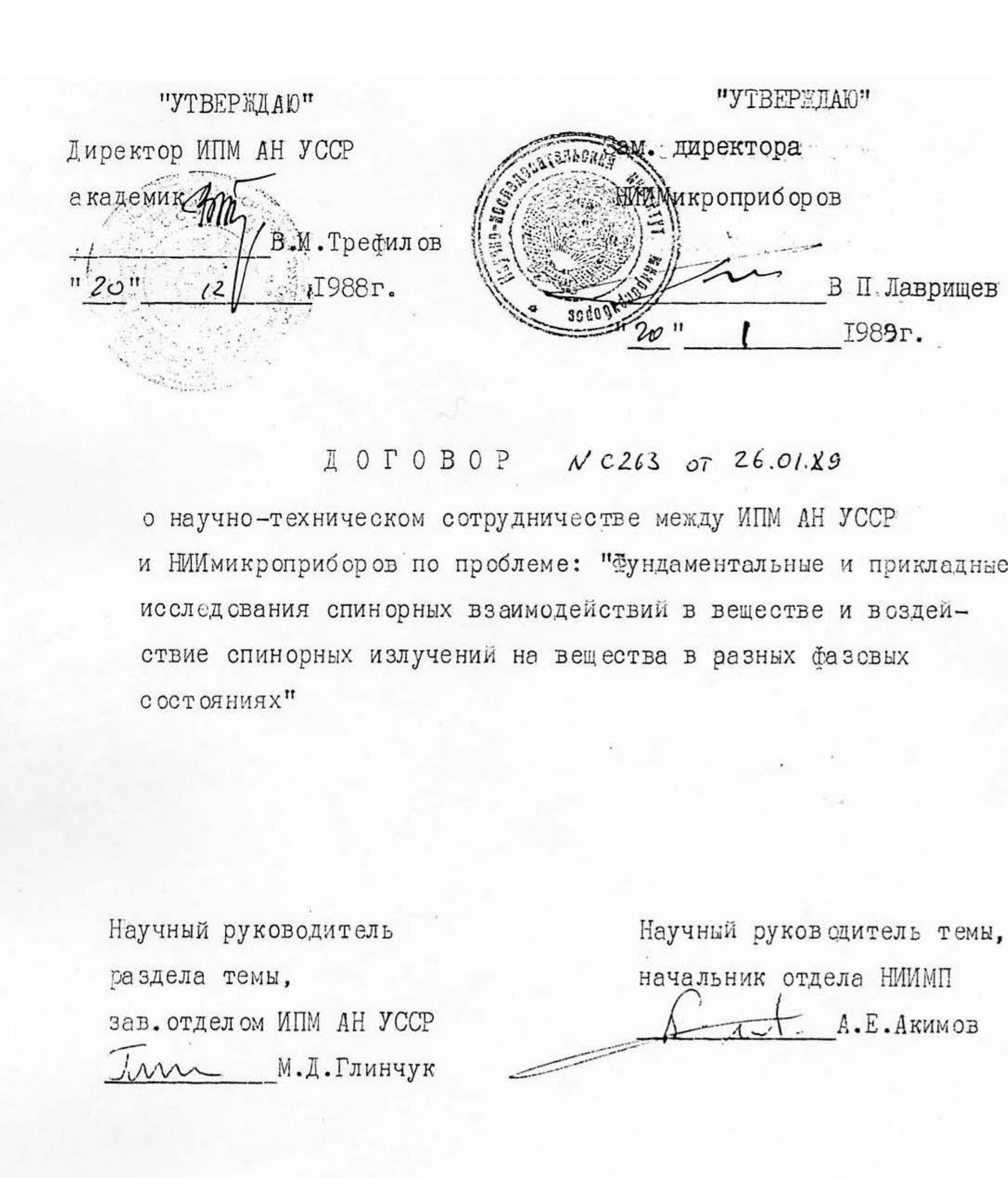}
\caption{\em \small Example of contracts between NII MP and other organizations for the study of 'spinor' emission of late 80s, from \cite{Zhigalov09}.
\label{fig:vertrag}}
\end{figure}

There are not many known facts about the appearance of the center 'Vent' and backstair financial deals within SCST USSR in 1986-1991. According to akimovae.com, A.E.Akimov in 1977-1983 worked at the Moscow Scientific Research Institute of Radio, in 1983-1987 at the Research Institute for Communication Systems and Control, and in 1987-1991 as the department head at the Research Institute of Microdevices (NII MP). Research topics covered by Akimov related to communication systems, thus we find the theme of communication as the point N1 in the list of unconventional technologies. There are many copies of contracts between the NII MP and the various research institutes regarding the study of the 'spin' fields, as shown in Fig.~\ref{fig:vertrag}, where NII MP provided generators of 'torsion fields' \cite{Dulnev04}. In these contracts A.E.Akimov acted as a supervisor on this topic. It can be concluded that the instrumental generators were developed by NII MP from the second half of the 80s, and the texts of resolutions for SCST USSR were prepared in this organization. However, the communication experiments from 1986 were performed at the Research Institute for Communication Systems and Control -- this implies that many top Soviet research institutes were involved in these programs. As mentioned in \cite{Vinokurov93}, the general Hantseverov also prepared a proposal for the program, however due to some reasons, Akimov's programme was accepted.

The publications of the center 'Vent' indicate a very good knowledge of corresponding literary sources, including Western patents, up to the beginning of the 20th century. For example, N4 in the list of technologies -- water cleaning with spin generators -- this topic has been developed in the 70s by R.Pavlita (unfortunately this topic was not further developed within 'Vent' activities), direct links to Western patents in the 'policy document' \cite{Akimov91}; wording of the patent \cite{AkimovPatent92}. On the one hand, it can be assumed that 'spinor' topics were carefully studied in the NII MP, probably long before 1988. On the other hand, such concepts as the microlepton field of A.F.Ohatrin, the axion field of V.Tatur (В.Татур) were not included into the 'Vent' program -- this points to a certain ideological competition\footnote{See e.g. the work of L.I.Holod, I.V.Goryachev, 'On vacuum models developed by Y.Terletskiy, G.Shipov, A.Akimov and A.Okhatrin-B.Tatur'.}. 'Old' works on morphogenetic fields of A.G.Gurvich (А.Г.Гурвич) or on 'psichons' of B.Kobozev were also not followed further.

The word 'torsion' appeared after 1990 (probably in 1991). Before 1990, the 'spin concept' was used. In the list  of Akimov's publications, one of the first works about spin systems refers to 1987 \cite{Akimov87}, in collaboration with L.B.Boldyreva and N.B.Sotina (Л.Б.Болдырева, Н.Б.Сотина). L.B.Boldyreva in \cite{Boldureva09} described that Akimov, around 1986-1987, was greatly impressed by the work A.A.Deev had done with spin-polarized materials. The basis of work \cite{Akimov87} represents the model of Dirac vacuum. After the start of cooperation with G.I.Shipov the concept received the name 'torsion' (or spin-torsion) and includes the torsion effects \cite{Shipov93}. However, analysing literature, e.g. \cite{Collins86} (prepared probably by intelligence service), comparing Puthoff's \cite{Puthoff98} and Akimov's patent \cite{AkimovPatent92}, we found a number of evidences, that the issue of 'torsion generators' is closely related to the Aharonov-Bohm effect \cite{Aharonov59}, and to several effects of quantum phenomena in macroscopic systems \cite{Vedral:2008p438}. It is very characteristic, that patents \cite{Puthoff98}, \cite{AkimovPatent92} and the work \cite{Collins86} appeared in a short time at the end of the 80s--beginning of 90s. Moreover, it seems this represents a connection between Russian 'spin-torsion' and Western quantum works.

Almost all experiments in the late 80s and early 90s within state programs were performed with so-called small and large Akimov's generator, see Fig.~\ref{fig:aktGen}.
Other generators of that time, e.g. developed by A.Y.Smirnov \cite{Smirnov10}, also had a similar philosophy for their design and operating principles. Later, there appeared other generators, based on the principles described in \cite{Akimov91}, and some new principles, such as LED generators of A.V.Bobrov \cite{bobrov01NEM}. An overview can be found in \cite{Zhigalov11}.

\textbf{3) Development of instrumental sensors.} One of the main problems associated with the 'high-penetrating' radiation is the problem of how to detect this radiation. Without generators it is not possible to develop the sensors, and without sensors it is not possible to design generators. This is well-known cycle of instrumental psychotronics, without resolving this problem it is impossible to do research in this area.

In the early 80's there were already some principles of detection known  based on Kozyrev's work (changing the conductivity of materials and the impact on mechanical systems), works of Chizhevskij (settling red blood cells) and the Kirlian effect (glow in the strong EM field). In addition to this, widely used methods were dowsing (more generally biolocation) and biological sensors, such as the germination test. In 1982-1984 some research on the Kozyrev's method was carried out, in particular with biological systems \cite{Danchakov84}. However, these works were fragmented, researchers often did not know each other. This situation substantially changed after 1989.

In 1988 the first note of A.V.Bobrov about the electric double layer (EDL) \cite{Bobrov88} as a sensor had appeared. In his book \cite{Bobrov06}, he confirmed the program of the USSR's Ministry of Defense on the study of psychics and instrumental psychotronics. Bobrov's EDL sensors have been proven as very sensitive devices. The book edited by Lunev \cite{Lunev95} describes the work carried out at the Tomsk Polytechnic University from 1983 to 1993, including a number of sensors based on quartz resonators and detectors of radioactivity. In 1989 a patent of G.A.Sergeev \cite{Sergeev89} on capacitive sensors is issued. Since 1989 various tests with crystallized structures are conducted  \cite{Akimov91B}, \cite{Panov02} and, further, with the melting of metals. Attempts were made to develop sensors on that basis. In the early 90s sensors of Y.P.Kravchenko appeared (Ю.П.Кравченко) \cite{Kravchenko94}, based on measurements of electric fields. In the Institute of Physics, St. Petersburg State University, the results related to Kozyrev's sensors are verified \cite{Bakaleinikov92} (these and other works have stimulated the development of solid-state sensors). The book of G.N.Dulnev and colleagues \cite{Dulnev98} describes the research conducted between 1995-1998 at the Center for Energy and Information Technologies at the St. Petersburg State Institute of Fine Mechanics and Optics (TSEIT GITMO). In those experiments more biological, optical, magnetic and thermal sensors are used. Interesting works are performed on bioelectrogenesis of plants \cite{Opritov91} and the application of such sensors in experiments \cite{Akimov01}. By 2000, there is already a large amount of works on the impact of 'high-penetrating' radiation on different semiconductor devices, see e.g. \cite{Asheulov00}. In the review \cite{Kernbach13metrology} in 2013 related to metrology of 'high-penetrating' emission, there are 19 groups of physical effects that can represent a basis for the development of sensors, with dozens of technical sensors.

The appearance of generators and detectors of 'high-penetrating' radiation in 2003-2004 indicated the beginning of a new stage in the development of instrumental psychotronics. These effects can by examined by independent researchers and even by non-technical (e.g. biologists) experts. This factor played a major role in the evolution of this research area after minimizing and closing governmental programs.

\section{Conclusion}
\label{sec:conclusion}

To sum up this review, first of all, we note a characteristic feature of Soviet and Russian programs -- specific position of the state, which during the entire existence of the Soviet Union has funded certain areas of parapsychology and blocked all others. In addition, since 1917, international contacts have been significantly reduced or even discontinued. As a result of this the USSR, since the early 70s, began to shape a specific area of instrumental psychotronics. By 2003-2004, with the development of sensors capable of detecting a 'high-penetrating' emission, an entirely independent line of research had emerged in Russia. This combined sources and detectors of radiation, and explored the appeared effects in a variety of materials and systems. Analyzing Western works, such as USPA\footnote{USPA - US Psychotronics Association, www.psychotronics.org} and SSE\footnote{The Society for Scientific Exploration (SSE), www.scientificexploration.org} conferences, the corresponding European and American magazines, we note only a small number of publications on related topics.

Second, an existence of many enthusiastic researchers in Russia after 1991 should be underlined. The concept of 'firm in the garage', widely known in the West, has a special meaning in Russia, since many studies have been performed not only by the academic community, but also by private researches. Regarding the difficulties in these works we can refer to the statements of Plekhanov \cite{Plexanov04}, Kulagina \cite{PhenomenD91}, Dulnev \cite{Dulnev92} Bobrov \cite{Bobrov06}, and many others. It should be noted that the Soviet Union coordinated the activities in this area at the Scientific Council of the USSR's Academy of Science in the 60s, 70s and 80s, while after 1991 the Academy of Sciences suddenly announced this research as 'pseudoscientific'. Such rapid change is surprising, and is an argument in favor of the hypothesis that this was primarily related to financial issues within the SCST USSR.

Regarding the funding we found the following data\footnote{Exchange rate of 1 Soviet ruble was slightly larger than 1USD.}: the DIA Report 'Controlled Offensive Behaviour USSR, July 1972' estimated the budget \$21 million in 1967 (in the area 1), Hantseverov's estimation \cite{May93RUS} about 700 involved researchers after 1987 (about 7 million ruble per year in the area 1), Akimov's estimation about the required 500 million ruble in 1990-1995 \cite{Zhigalov09} (the area 3), data from \cite{Zhigalov09} about spent 23 million ruble from Ministry of Defence around 1986-1989, data from \cite{May93RUS} with estimated 400 involved researchers after 1990 (about 4 million ruble per year in the area 1), data from \cite{Ptichkin09} about the military unit 10003, which consumed 4 million ruble per year and existed for 15 years (the area 1). Based on these data it can be estimated that open funding in the 80s and 90s was between \$200 million and \$400 million. Extrapolating data from the 60s and 70s it can be assumed that the upper boundary for open funding in after-war programs was about \$500 million.

However, there is no data about studies from KGB and USSR's Ministry of Defense in areas 1 and 2. A.V.Bobrov in \cite{Zhigalov09} and in discussions reported about studies performed at Gagarin Air Force Academy in Monino and by  St. Petersburg I.I.Mechnikov State Medical Academy in the 80s and 90s. These studies are open and covered a large number of technical systems and human operators. It would make sense to assume that classified programs in the area 1 were larger than open programs. Since the area 2 was a strategic program, its funding should be assumed larger than of 1 and 3. In total we can assume an upper  boundary of \$0.5-1 billion for all areas within 40 years. It is vital to repeat these are estimates based on published data and some common sense consideration. To compare, U.S. Startgate program costs about \$20 million \cite{Waller95}, some programs for nonlethal technologies \$37.2 million \cite{Brandt96}, MKULTRA -- \$87.5 million \cite{Brandt96}. Thus, Soviet and U.S. costs are comparable at least on a level of separate programs.

Third, it needs to underline the ethical position in instrumental psychotronics. A.E.Akimov in 1995 at the conference 'KGB: Yesterday, Today, Tomorrow' wrote:
\begin{quote}
'As we are the leading organization in Russia on torsion technologies, I can responsibly say that while technological experiments allow sometimes obtaining results that go far beyond our fantasy, modern torsion generators are rather primitive, and it is difficult to expect that tomorrow or the day after tomorrow may appear sources of torsion radiation that could solve the problem of controlling human behavior. I do not believe in this. There is an international organization of scientists involved in the study of electromagnetic radiation, including their effect on humans. From a technical point of view, there is no reason why it would be impossible to make a machine, which impacts the human. And I have no doubt about the fact that this kind of technique exists in many countries of the world' \cite{Glasnist95}.
\end{quote}

The technology that was primitive in 1995, has become less primitive in 2013, and will be even more advanced in the future. Even now it is not too difficult to develop a high-power generator based on e.g. the Puthoff's patent \cite{Puthoff98}, Akimov's patent \cite{AkimovPatent92} or results of many other researchers, e.g. \cite{Shkatov01}. We also do not believe that human behavior can be controlled. However, we want to draw attention to the significant potential of a long-term use of these devices and the risk of unethical use of this technology for so-called 'mild correction', such as the PID effect\footnote{\textbf{P}erenos \textbf{I}nfomacionnogo \textbf{D}ejstvia -- transfer of information actions.} studied in plants and laboratory animals, see \cite{maslobrod12}, \cite{maslobrod09}, \cite{Maslobrod13}, \cite{maslobrod11}, \cite{Smirnov10}, \cite{Smirnov}, \cite{Krasnobrygev09}, \cite{Novikov13} and others.

To conclude this work, it should be noted that after 2003, the instrumental psychotronics in Russia was developing further, although not as fast as before. There are conferences, workshops, seminars and published specialized journals. According to \cite{May93RUS}, the number of Russian researchers in the areas around instrumental psychotronics is between 400 and 700 between 1987-1993. Currently, based on major conferences, this number can be estimated between 200 and 500. We believe that this figure is kept at this level and beyond, but there is a noticeable increase in the age of researchers. There is also an interesting process of distributing this research to the west. First of all, it is done by researchers from CIS counties and who is living in Western countries. This is also happening due to the inclusion of countries from the former Soviet Union and the Warsaw Pact in the EU and the gradual introduction of these works in Western research organizations.

\section{Acknowledgment}

Author would like to thank the group 'Second physics' and Association of Unconventional Science for fruitful discussions and multiple comments for improving this work. Especial thanks are expressed to A.V.Bobrov and V.A.Zhigalov for discussions about historical evolution of unconventional research in Russia.

\small
\renewcommand{\refname}{References (Original Language)}

\renewcommand{\refname}{References (Translit)}

\end{document}